\documentclass[twocolumn,showpacs,showkeys,preprintnumbers,superscriptaddress,amsmath,floatfix,amssymb,prd]{revtex4}
\usepackage[colorlinks=true]{hyperref}
\usepackage{graphicx}

\newtheorem{alg}{Algorithm}

\newcommand{\comment}[1]{}

\newcommand{\lr}[1]{ \left( #1 \right) }
\newcommand{\lrs}[1]{ \left[ #1 \right] }
\newcommand{\lrc}[1]{ \left\{ #1 \right\} }
\newcommand{\vev}[1]{ \langle \, #1 \, \rangle }

\newcommand{\erfc}[1]{ {\rm Erfc}\left( #1 \right) }

\newcommand{\expa}[1]{ \exp{\left( #1 \right)} }

\newcommand{\logo}{\\ \vskip -18mm
\leftline{\includegraphics[scale=0.3,clip=false]{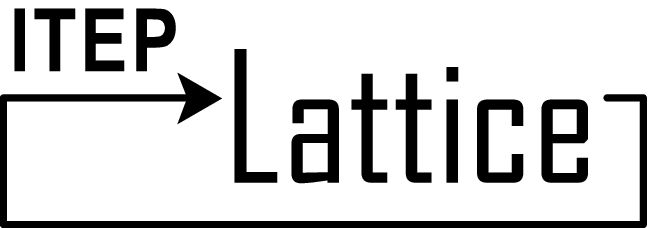}} \vskip 10mm}

\begin{document}
\sloppy
\preprint{ITEP-LAT/2011-05}

\title{A method for resummation of perturbative series based on the stochastic solution of Schwinger-Dyson equations\logo}

\author{P. V. Buividovich}
\email{buividovich@itep.ru}
\affiliation{ITEP, 117218 Russia, Moscow, B. Cheremushkinskaya str. 25}
\affiliation{JINR, 141980 Russia, Moscow region, Dubna, Joliot-Curie str. 6}

\date{September 16, 2011}
\begin{abstract}
 We propose a numerical method for resummation of perturbative series, which is based on the stochastic perturbative solution of Schwinger-Dyson equations. The method stochastically estimates the coefficients of perturbative series, and incorporates Borel resummation in a natural way. Similarly to the ``worm'' algorithm, the method samples open Feynman diagrams, but with an arbitrary number of external legs. As a test of our numerical algorithm, we study the scale dependence of the renormalized coupling constant in a theory of one-component scalar field with quartic interaction. We confirm the triviality of this theory in four and five space-time dimensions, and the instability of the trivial fixed point in three dimensions.
\end{abstract}
\maketitle

\section*{Introduction}
\label{sec:introduction}

 One of the most popular methods for numerical simulation of quantum field theories is the Monte-Carlo integration over all possible field configurations, with the integration weight being proportional to $\expa{-S}$, where $S$ is the action of the theory. However, such method might become not very efficient when the fields are strongly correlated at large distances, for example, in the vicinity of quantum phase transitions, or when the path-integral weight becomes non-positive. The latter situation is typical, e.g., for field theories at finite chemical potential.

 Recently it has been realized that these difficulties can be avoided or significantly reduced if one performs the perturbative expansion of the theory around some ``free'' action $S_0$, and sums up the resulting series, instead of directly evaluating the path integral. Such summation can be also performed using the Monte-Carlo procedure, and this alternative simulation method is usually called ``Diagrammatic Monte-Carlo'' \cite{Prokofev:98:1, Wolff:09:1}. Diagrammatic Monte-Carlo turned out to be very efficient for numerous problems in statistical and condensed matter physics, and allowed to obtain many interesting results with precision which was unattainable for other methods.

 However, a systematic application of Diagrammatic Monte-Carlo to field theories which are relevant for high-energy physics is so far hindered mainly because expansions in powers of coupling constant, which lead to the conventional diagrammatic technique due to Feynman, typically yield only asymptotic series. Such series cannot be directly summed and therefore cannot be sampled by a Monte-Carlo procedure. In practical simulations, this non-convergence problem is typically avoided by finding a suitable strong-coupling expansion. It is then argued that in a finite volume this expansion can be continued to the weak-coupling domain \cite{Wolff:09:1}. Examples of such expansions which were used for numerical simulations are the strong-coupling expansion in $O\lr{N}$ and $CP_{N}$ lattice sigma-models and in Abelian gauge theories, and the Aizenman random current representation \cite{Aizenman:81:1} for the $\phi^4$ theory \cite{Wolff:09:1}. Unfortunately, the structure of strong-coupling expansions differs significantly for different theories. Moreover, in many cases they are quite complicated and their structure is not sufficiently well understood. In particular, it has not been realized yet how to systematically sample strong-coupling expansion diagrams for non-Abelian lattice gauge theories or for sigma-models with $SU\lr{N}$ target space. Some progress in this direction has been recently achieved for field theories in the large-$N$ limit \cite{Buividovich:10:2}.

 Another way to treat the non-convergent weak-coupling expansions has been proposed recently in \cite{Prokofev:10:1}. The basic idea is to construct a sequence of better and better approximations to the original path integral, each of which has a convergent expansion. However, so far the utility of this construction was demonstrated only for the zero-dimensional theory. It is also not clear how to generalize the construction of \cite{Prokofev:10:1} to lattice field theories with compact variables, such as non-Abelian lattice gauge theories. One can conclude that in the context of numerical simulations in high-energy physics Diagrammatic Monte Carlo is so far a promising, but not a universal tool.

 In the present paper we propose a novel simulation method which, on the one hand, inherits the advantageous features of Diagrammatic Monte-Carlo, and, on the other hand, can be used to investigate theories with asymptotic weak-coupling expansions and does not require the detailed knowledge of the structure of the perturbative expansion. The method is based on the stochastic interpretation of Schwinger-Dyson equations, which are typically much easier to derive than the general form of the coefficients of perturbative series. Such stochastic interpretation has been considered recently in \cite{Buividovich:10:2} for large-$N$ quantum field theories, and a somewhat similar approach was discussed quite a long time ago in \cite{Marchesini:81:1} (see \cite{Buividovich:10:2} for a more detailed discussion of the methods used in these papers). Similarly to the ``worm algorithm'' \cite{Prokofev:98:1}, which is an essential ingredient in the Diagrammatic Monte-Carlo, the method samples open diagrams which correspond to field correlators, rather than closed diagrams which correspond to the partition function of the theory. Another distinct feature is that the diagrams are sampled directly in the momentum space. Asymptotic series can be treated within this method using the standard resummation tools, such as the Pade-Borel resummation or expansion over the basis of some special functions \cite{KleinertPhi4}. For example, in the case of factorially divergent series, our method incorporates the Borel transform (or, more generally, the Borel-Leroy transform) of the field correlators in a natural way.

 In order to illustrate the applicability of our method, here we consider the theory in which Schwinger-Dyson equations take probably the simplest nontrivial form, namely, the theory of a one-component scalar field with quartic interaction. Perturbative series in this theory are known to diverge factorially, and we use Pade-Borel resummation to recover physical results. As a test of the method, we study the scale dependence of the renormalized coupling constant of the theory in different space-time dimensions, and confirm numerically the triviality of the theory in five and four space-time dimensions and the instability of the trivial fixed point in three dimensions \cite{Aizenman:81:1, Frohlich:82:1}. For simplicity, we work only in the phase with unbroken $Z_2$ symmetry. Application of our algorithm to the phase with broken symmetry will be discussed briefly in the concluding Section, and will be studied separately elsewhere. We hope that the simulation strategy, which we illustrate here on the simplest nontrivial example, can be easily generalized to other quantum field theories.

 The paper is organized as follows: in Section \ref{sec:SDs_phi4} we introduce the basic definitions and write down the Schwinger-Dyson equations for the $\phi^4$ theory in momentum space. In Section \ref{sec:lineq_stochastic} we provide a general description of the stochastic method which we use to solve these equations. Then in Section \ref{sec:stochastic_sd_solution} we formulate the simulation algorithm which stochastically estimates the coefficients of the perturbative expansion of the field correlators in powers of the bare coupling constant. In Section \ref{sec:series_resummation} we consider a practical method to extract physical observables from the numerical data produced by this algorithm. While Sections \ref{sec:SDs_phi4} - \ref{sec:stochastic_sd_solution} are essential to understand the presented method, Section \ref{sec:series_resummation} is more technical. In Section \ref{sec:phys_res} we present and briefly discuss some physical results which we obtained using our algorithm. Finally, in Section \ref{sec:conclusions} we make some concluding remarks and discuss further extensions and generalizations of our approach.

\section{Schwinger-Dyson equations for a single-component $\phi^4$ theory}
\label{sec:SDs_phi4}

 We consider the theory of a one-component scalar field $\phi\lr{x}$ in $D$-dimensional Euclidean space with quartic interaction. The action of the theory is:
\begin{eqnarray}
\label{phi4_action}
 S = \int d^D x \lr{
 1/2 \, \partial_{\mu} \phi \partial_{\mu} \phi +
 m_0^2/2 \, \phi^2 + \lambda_0/4 \, \phi^4
 }  .
\end{eqnarray}
The disconnected field correlators in momentum space are defined as follows:
\begin{eqnarray}
\label{phi4_correlators_def}
\phi\lr{p} = \int d^D x \, e^{i p x} \, \phi\lr{x}
\nonumber \\
G\lr{p_1, \ldots, p_n} = \vev{\phi\lr{p_1} \ldots \phi\lr{p_n}}  .
\end{eqnarray}
Note that only even-order correlators are nonzero. By $G_c\lr{p_1, \ldots, p_n}$ we denote the corresponding connected correlators (that is, the correlators which contain only connected Feynman diagrams). Due to momentum conservation, all correlators also contain a factor $\lr{2 \pi}^D \, \delta\lr{\sum \limits_{A = 1}^{n} p_A }$. It is also convenient to define the correlators $G'\lr{p_1, \ldots, p_n}$ from which this factor is omitted.

 We define the renormalized mass $m_R$ and the wave function renormalization constant $Z_R$ from the behavior of the two-point correlator at small momenta \cite{Luscher:87:1}:
\begin{eqnarray}
\label{renorm_mass_def}
G'\lr{p_1, p_2} = \frac{Z_R}{p_1^2 + m_R^2 + O\lr{p_1^4}}
\end{eqnarray}
The renormalized coupling constant $\lambda_R$ is related to the one-particle irreducible four-point correlator $\Gamma\lr{p_1, p_2, p_3, p_4}$ at zero momenta. For the theory (\ref{phi4_action}) it is proportional to the connected four-point correlator with truncated external legs:
\begin{eqnarray}
\label{renorm_coupling_def}
\lambda_R = - 1/6 \, Z_R^2 \, \Gamma\lr{0, 0, 0, 0}
 = \nonumber \\ = 
- 1/6 \, Z_R^2 \, \lr{G'\lr{0,0}}^{-4} \, G_c'\lr{0, 0, 0, 0}
\end{eqnarray}
We note that our definition of the coupling constant differs from the one that is used in most textbooks by a factor of $6$. This definition is more convenient for the analysis of Schwinger-Dyson equations.

 Schwinger-Dyson equations for the correlators (\ref{phi4_correlators_def}) in momentum space read:
\begin{widetext}
\begin{eqnarray}
\label{phi4_sd_disconnected_n2}
\lr{m_0^2 + p_1^2} \, G\lr{p_1, p_2} = \lr{2 \pi}^D \delta\lr{p_1 + p_2}
 - \nonumber \\ -
\frac{\lambda_0}{\lr{2 \pi}^{2 D}}
\int d^D q_1 \, d^D q_2 \, d^D q_3
\delta\lr{p_1 - q_1 - q_2 - q_3} G\lr{q_1, q_2, q_3, p_2}
\end{eqnarray}
\end{widetext}
\begin{widetext}
\begin{eqnarray}
\label{phi4_sd_disconnected}
\lr{m_0^2 + p_1^2} \, G\lr{p_1, p_2, \ldots, p_n}
 =
\sum \limits_{A=2}^{n} \lr{2 \pi}^D \delta\lr{p_1 + p_A}
G\lr{p_2, \ldots, p_{A-1}, p_{A+1}, \ldots, p_{n}}
 - \nonumber \\ -
\frac{\lambda_0}{\lr{2 \pi}^{2 D}}
\int d^D q_1 \, d^D q_2 \, d^D q_3 \,
\delta\lr{p_1 - q_1 - q_2 - q_3} \, G\lr{q_1, q_2, q_3, p_2, \ldots, p_n}  ,
\end{eqnarray}
\end{widetext}
where the arguments of the correlator in the first summand on the r.h.s. of (\ref{phi4_sd_disconnected}) are all the momenta except $p_1$ and $p_A$.

 Equations (\ref{phi4_sd_disconnected_n2}) and (\ref{phi4_sd_disconnected}) were obtained by variation of the field correlators and the action over $\phi\lr{p_1}$. Similar equations can be obtained for any argument of field correlators $p_1, \ldots, p_n$, but the resulting system of equations turns out to be redundant. To obtain a complete system of equations, it is sufficient to consider the variation over $\phi\lr{p_1}$ only. We thus arrive at a system of functional linear inhomogeneous equations for an infinite set of unknown functions $G\lr{p_1, \ldots, p_n}$. According to the general theorems of linear algebra, the solution of such equations is unique, if it exists. A straightforward way to solve equations (\ref{phi4_sd_disconnected_n2}), (\ref{phi4_sd_disconnected}) is to truncate an infinite set of equations at some correlator order and to discretize the continuum momenta. In this case, however, the required computational resources quickly grow with the number of correlators which are studied. Such infinite systems of linear equations can be more efficiently solved using stochastic methods. In the next Section we provide a general description of a specific stochastic method which, in our opinion, is most convenient for the solution of Schwinger-Dyson equations in quantum field theories.

\section{Stochastic solution of inhomogeneous linear equations}
\label{sec:lineq_stochastic}

 We consider a system of linear inhomogeneous equations of the following form:
\begin{eqnarray}
\label{lineq0}
 f\lr{x} = \sum\limits_{y \in X} A\lr{x, y} F\lr{x, y} f\lr{y} + b\lr{x}
\end{eqnarray}
where $x$ is the element of some space $X$ and $\sum \limits_{x \in X}$ denotes summation or integration over all elements of this space. The coefficients $A\lr{x, y}$ are assumed to be positive, while $F\lr{x, y}$ can be of any sign. $A\lr{x, y}$ and $F\lr{x, y}$ are also assumed to satisfy the inequalities
\begin{eqnarray}
\label{lineq_ineq}
 \sum \limits_{x \in X} A\lr{x, y} < 1
\nonumber \\
 |F\lr{x, y}| < 1
\end{eqnarray}
for any $x, y \in X$. The source function $b\lr{x}$ is also assumed to be positive. Factorization of the coefficients of the equation (\ref{lineq0}) into $A\lr{x, y}$ and $F\lr{x, y}$ is to a large extent arbitrary, and can be chosen in some optimal way for any particular problem.

 If the space $X$ in (\ref{lineq_ineq}) contains an infinite number of elements, a deterministic approximate solution of (\ref{lineq0}) would require the truncation of this space to some finite number of elements, and it is \emph{a priori} not known which elements can be discarded. An alternative to the deterministic solution is the stochastic solution, for which $f\lr{x}$ is proportional to the probability of occurence of the element $x$ in some random process. This idea dates back to the works of von Neumann and Ulam (described in \cite{Forsythe:1950:1}, see also \cite{Srinivasan:03:1} for a more recent review). In this case, the solution automatically incorporates the importance sampling. Namely, those elements of space $X$ for which $f\lr{x}$ is numerically sufficiently small are automatically truncated, since the random process cannot reach them in a finite number of iterations. Such methods for the solution of large systems of linear equations are widely used, e.g., in engineering and control design, for problems related to partitions of large graphs etc. Since the basic idea behind these methods is very simple, and the number of works on the subject is vast, we have decided to present here a concise formulation of the method which we found most useful, without giving any specific reference to the literature.

 Consider the Markov process with states specified by one element of the space $X$ and one real number $\chi$. We specify this process by the following
\begin{alg}
\label{alg:rp_abstract_desc}
At each iteration, do one of the following:
\begin{description}
  \item[Evolve:] With probability $A\lr{x, y}$ change from the current state $y$ to the new state $x$ and multiply $\chi$ by $F\lr{x, y}$.
  \item[Restart:] Otherwise go to the state with $\chi=1$ and a random $z \in X$, which is distributed with probability $b\lr{z}/\lr{\sum \limits_{x} b\lr{x}}$.
\end{description}
\end{alg}
The ``Restart'' action is also used to initiate the random process. The first inequality in (\ref{lineq_ineq}) ensures that the probability of the ``Evolve'' action does not exceed unity, and the second inequality ensures that $\chi$ remains bounded and thus have a well-defined average over the stationary state of the Markov process.

 Now let $w\lr{x, \chi}$ be the stationary probability distribution for this random process, that is, the probability to find the process in the state $\lrc{x, \chi}$ measured over a sufficiently large number of iterations of Algorithm \ref{alg:rp_abstract_desc}. A general form of the equation governing the stationary probability distributions of Markov processes is $w\lr{A} = \sum \limits_{B} P\lr{B \rightarrow A} w\lr{B}$, where $P\lr{B \rightarrow A}$ is the probability of transition from the state $B$ to the state $A$. For the random process specified above such an equation reads:
\begin{widetext}
\begin{eqnarray}
\label{lineq_markov}
 w\lr{x, \chi} = \sum\limits_{y \in X}  A\lr{x, y} \, \int d\chi' \, \delta\lr{\chi, F\lr{x, y} \chi'} \, w\lr{y, \chi'}
+ \nonumber \\ +
\frac{b\lr{x}}{\lr{\sum \limits_{z} b\lr{z}}} \, \delta\lr{\chi, 1} \, \sum\limits_{y} \int d\chi' \, w\lr{y, \chi'} \, \lr{1 - \sum\limits_{z \in X} A\lr{z, y}}
\end{eqnarray}
\end{widetext}
We note that $w_R = \sum\limits_{y} \int d\chi' \, w\lr{y, \chi'} \, \lr{1 - \sum\limits_{z \in X} A\lr{z, y}}$ is the average probability of choosing the ``Restart'' action. Let us denote $\mathcal{N}^{-1} = w_R/\lr{\sum \limits_{z} b\lr{z}}$. By integrating (\ref{lineq_markov}) over $\chi$ one can check that
\begin{eqnarray}
\label{lineq_solution}
f\lr{x} = \mathcal{N} \int d\chi \, \chi \, w\lr{x, \chi}
\end{eqnarray}
is the solution of the original equation (\ref{lineq0}).

 In practice, in order to find $f\lr{x}$ one should simulate the random process specified by the Algorithm \ref{alg:rp_abstract_desc}, and sum up the factors $\chi$ over a sufficiently large number of iterations separately for each $x$, dividing the results by a total number of iterations:
\begin{eqnarray}
\label{averaging_formula}
f\lr{x} = \frac{\sum\limits_{z} b\lr{z}}{w_R} \,
\lim\limits_{T \rightarrow \infty} \frac{1}{T} \, \sum \limits_{t = 1}^{T}
\delta\lr{x, x_t} \chi_t ,
\end{eqnarray}
where $x_t$ and $\chi_t$ are the values of $x$ and $\chi$ at $t$'th iteration. Here $w_R$ can be measured simultaneously with $f\lr{x}$ as $w_R = \lim\limits_{T \rightarrow \infty} n_R/T$, where $n_R$ is the number of ``Restart'' actions during $T$ iterations.

\section{Stochastic interpretation of Schwinger-Dyson equations}
\label{sec:stochastic_sd_solution}

 At a first sight, one can try to apply the method described in Section \ref{sec:lineq_stochastic} directly to equations (\ref{phi4_sd_disconnected_n2}) and (\ref{phi4_sd_disconnected}). The space $X$ should be then the space of sequences of momenta $\lrc{p_1, \ldots, p_n}$ for any $n = 2, 4, \ldots$. However, a simple analysis shows that the inequalities (\ref{lineq_ineq}) for equations (\ref{phi4_sd_disconnected_n2}) and (\ref{phi4_sd_disconnected}) cannot be satisfied, since the number of summands in the first term on the r.h.s. of (\ref{phi4_sd_disconnected}) grows linearly with the number $n$ of field variables in the correlator. As will become clear from what follows, this growth is in fact the manifestation of the factorial divergence of the coefficients of the perturbative series. Such structure of Schwinger-Dyson equations for one-component scalar field should be contrasted with scalar field theories in the large-$N$ limit, where the number of planar diagrams grows only in geometric progression, and where Schwinger-Dyson equations in factorized form at sufficiently small coupling constant can be interpreted as equations for the stationary probability distributions of the so-called nonlinear random processes \cite{Buividovich:10:2}.

 In order to overcome this problem, let us assume that the correlators (\ref{phi4_correlators_def}) can be formally expanded in power series in the coupling constant $\lambda_0$:
\begin{eqnarray}
\label{correlators_prob_redef}
G\lr{p_1, \ldots, p_n} = \sum \limits_{m=0}^{+\infty} c_{n, m} \, \lr{-\lambda_0}^m \, G_m\lr{p_1, \ldots, p_n}
\end{eqnarray}\
where the coefficients $c_{n, m}$ will be specified later.

 Now we insert the expansion (\ref{correlators_prob_redef}) into the Schwinger-Dyson equations (\ref{phi4_sd_disconnected_n2}) and (\ref{phi4_sd_disconnected}) and collect the terms with different powers of $\lambda_0$. This yields the following equations for the functions $G_m\lr{p_1, \ldots, p_n}$:
\begin{widetext}
\begin{eqnarray}
\label{stochastic_eq_n2}
G_m\lr{p_1, p_2} =
\frac{\delta_{m, 0} \, \lr{2 \pi}^D \delta\lr{p_1 + p_2}}{c_{2, 0} \, \lr{m_0^2 + p_1^2}}
 + \nonumber \\ +
\frac{c_{4, m-1}}{c_{2, m} \, \lr{2 \pi}^{2 D} \, \lr{m_0^2 + p_1^2}}
\int d^D q_1 \, d^D q_2 \, d^D q_3 \,
\delta\lr{p_1 - q_1 - q_2 - q_3} G_{m-1}\lr{q_1, q_2, q_3, p_2}
\end{eqnarray}
\end{widetext}
\begin{widetext}
\begin{eqnarray}
\label{stochastic_eq}
G_m\lr{p_1, \ldots, p_n} =
\sum \limits_{A=2}^{n} \frac{c_{n-2, m} \, \lr{2 \pi}^D \, \delta\lr{p_1 + p_A}}{c_{n, m} \, \lr{m_0^2 + p_1^2}}
G_m\lr{p_2, \ldots, p_{A-1}, p_{A+1}, \ldots, p_{n}}
 + \nonumber \\ +
\frac{c_{n+2, m-1}}{c_{n, m} \lr{2 \pi}^{2 D} \lr{m_0^2 + p_1^2}}
\int d^D q_1 d^D q_2 d^D q_3 \delta\lr{p_1 - q_1 - q_2 - q_3} G_{m-1}\lr{q_1, q_2, q_3, p_2, \ldots, p_n}
\end{eqnarray}
\end{widetext}

 Equations (\ref{stochastic_eq_n2}) and (\ref{stochastic_eq}) are also linear inhomogeneous equations, but on a larger functional space - the unknown functions $G_m\lr{p_1, \ldots, p_n}$ now depend also on $m$. We can now try to choose the coefficients $c_{n, m}$ in (\ref{correlators_prob_redef}) so as to cast these equations in the stochastic form (\ref{lineq0}) with coefficients which satisfy the inequalities (\ref{lineq_ineq}). The space $X$ now should contain sequences $\lrc{p_1, \ldots, p_n}$ of $n \ge 2$ momenta in $D$-dimensional space and a nonnegative integer number $m$. The sum $\sum \limits_{x \in X}$ should be understood as summation over $n$ and $m$ and integration over $n$ momenta:
\begin{eqnarray}
\label{summation_def}
\sum \limits_{x \in X} \rightarrow \sum \limits_{m \ge 0} \sum \limits_{n \ge 2} \int d^D p_1 \, \ldots d^D p_n
\end{eqnarray}
Thus, altogether the space of states of the Markov process that we would like to construct consists of an ordered sequence of momenta of arbitrary length $\lrc{p_1, \ldots, p_n}$, a positive integer number $m$ and a real number $\chi$.

 Comparing the form of equations (\ref{stochastic_eq_n2}) and (\ref{stochastic_eq}) with the general form (\ref{lineq0}), we can now construct Markov process which will solve these equations, as discussed in Section \ref{sec:lineq_stochastic}. Again, we specify it by the following
\begin{alg}
\label{alg:rp_phi4}
At each iteration, do one of the following:
\begin{description}
 \item[Add momenta:] With probability $p_{A} = \frac{\lr{2 \pi}^D \, \lr{n+1} \, c_{n, m} \Sigma_0}{c_{n+2, m}}$, where $\Sigma_0 = \int\limits_{|p|<\Lambda} \frac{d^D p}{p^2 + m_0^2}$, add a pair of momenta $\lrc{p, -p}$ to the current sequence of momenta, inserting the first momenta at the beginning of the sequence and the second - between the $A$'th and $A+1$'th elements of the sequence, where $A$ is chosen at random between $\lr{n + 1}$ possibilities. The momentum $p$ is distributed within the $D$-dimensional sphere of radius $\Lambda$ with the probability distribution $\sim \frac{1}{p^2 + m_0^2}$. Do not change $m$ and $\chi$.
 \item[Create vertex:] With probability $p_{V} = \frac{c_{n, m}}{\lr{2 \pi}^{2 D} \, c_{n-2, m+1} \, m_0^2}$ replace the three first momenta $p_1$, $p_2$ and $p_3$ in the sequence by their sum $p = p_1 + p_2 + p_3$. Multiply $\chi$ by $m_0^2/\lr{m_0^2 + p^2}$ and increase $m$ by one.
 \item[Restart:] Otherwise restart with a sequence which contains a pair of random momenta $\lrc{p, -p}$ with the probability distribution $\sim \frac{1}{p^2 + m_0^2}$, and with $\chi = 1$ and $m = 0$.
\end{description}
\end{alg}

 Since the factor $m_0^2/\lr{m_0^2 + p^2}$ in the ``Create vertex'' action does not exceed unity, the second inequality in (\ref{lineq_ineq}) is satisfied. Let us check whether we can also satisfy the first inequality, that is, whether the total probability of ``Add momenta'' and ``Create vertex'' actions can be made less than one for any sequence of momenta and for any $m$. This can only be achieved if both ratios $\frac{\lr{n+1} \, c_{n, m}}{c_{n+2, m}}$ and $\frac{c_{n, m}}{c_{n-2, m+1}}$ are finite for any $n$ and $m$. For the first ratio it is only possible if $c_{n, m}$ grows not slower than $\lr{n/2}!$ at large $n$. In this case the second ratio can only be bounded for all $n$ and $m$ if $c_{n, m}$ grows as $\lr{n/2 + m}!$ at large $n$, $m$. We thus conclude that equations (\ref{stochastic_eq_n2}) and (\ref{stochastic_eq}) indeed can be interpreted as the equations for the stationary probability distribution of some Markov process, if the coefficients $c_{n, m}$ grow sufficiently fast. The factorial divergence of the perturbative series is then absorbed in the coefficients $c_{n, m}$. It is interesting that such a simple analysis of Schwinger-Dyson equations reveals the divergence of the perturbative series in a straightforward way, without the need to explicitly calculate any diagrams!

 In Algorithm \ref{alg:rp_phi4} we have also introduced the ultraviolet cutoff $\Lambda$, which is necessary in order to normalize the probability distribution of the momenta which are created when the ``Add momenta'' action is chosen. In our simulations we set $\Lambda = 1$, so that all masses are measured in units of $\Lambda$ (``lattice units''). A more self-consistent way to introduce such cutoff would be probably to start with lattice theory in the coordinate space and assume that all momenta belong to the first Brillouin zone $-\pi \le p_{\mu} \le \pi$. According to the standard renormalization-group arguments, a particular choice of the cutoff prescription should result only in tiny corrections of order $O\lr{e^{-1/\lambda_0}}$ to the physical results. Therefore we use the isotropic cutoff scheme, which leads to a much more efficient numerical algorithm. Indeed, random momenta with isotropic probability distribution can be easily generated by the standard mapping of the one-dimensional probability distribution $w\lr{|p|} \sim \frac{|p|^{D-1}}{|p|^2 + m_0^2}$ to the uniform probability distribution on the interval $\lrs{0, 1}$. On the other hand, with lattice discretization in coordinate space the random momenta should be generated with the anisotropic probability distribution $w\lr{p_\mu} \sim \frac{1}{m_0^2 + \sum\limits_{\mu} \sin^2\lr{p_\mu/2} }$. This can be done either by using the Metropolis algorithm or by discretizing the momenta, which require much more computational resources. An interpretation of our cutoff scheme in terms of Feynman diagrams will be given below.

 Algorithm \ref{alg:rp_phi4} thus stochastically samples the coefficients of the perturbative expansion of field correlators of the theory (\ref{phi4_action}), which are reweighted by the factors $c_{n, m}$. Each state of the Markov process defined by this algorithm corresponds therefore to some (in general, disconnected) Feynman diagram with $n$ external legs and $m$ vertices. Combinatorial growth of the number of diagrams is compensated by the growth of the coefficients $c_{n, m}$. The action ``Add momenta'' corresponds to adding a bare propagator line to the diagram, without attaching its legs anywhere - thus the number of external legs is increased by two, and the diagram order is not changed. The action ``Create vertex'' corresponds to the creation of one more vertex by joining three external legs of the diagram and attaching a new external leg to the joint. The number of external legs is thus decreased by two and the order of the diagram is increased by one. Since we add new momenta in pairs $\lrc{p, -p}$, the sum of momenta on all the external legs of any generated diagram is always identically zero. The constraint $\sum \limits_{A = 1}^{n} p_A = 0$ is thus automatically satisfied.

 By a slight modification of Algorithm \ref{alg:rp_phi4}, one can also easily trace whether the generated diagram is connected or disconnected. Upon the ``Add momenta'' action one should mark the two newly created legs by some unique label, and upon the ``Create vertex'' action, the labels of the legs which are being joined should be replaced by a single new label. In this way, one can directly measure the four-point connected diagram which enters the definition of the renormalized coupling constant (\ref{renorm_coupling_def}).

 The ``Restart'' action simply erases any diagram which was constructed before, and initiates the construction of a new diagram, which starts with just a single bare propagator. Since the maximal achievable diagram order is equal to the maximal number of ``Create vertex'' actions between the two consecutive ``Restart'' actions, it is advantageous to minimize the probability of ``Restarts''. The rate of ``Restart'' actions $w_R$ for our Algorithm \ref{alg:rp_phi4} thus plays the role similar to the reject probability in the standard Metropolis-based Monte-Carlo: for maximal numerical efficiency it is advantageous to maximally reduce it. On the other hand, in order to implement importance sampling with maximal efficiency, the factor $\chi$ in Algorithms \ref{alg:rp_abstract_desc} and \ref{alg:rp_phi4} should be as close to unity as possible. For each particular implementation of Algorithm \ref{alg:rp_abstract_desc}, one should find optimal balance between the rate of ``Restart'' actions and the efficiency of importance sampling by adjusting the factors $F\lr{x, y}$ in (\ref{lineq0}).

 In Algorithm \ref{alg:rp_phi4} we have already chosen these factors to be $m_0^2/\lr{m_0^2 + p^2}$ for the ``Create vertex'' action, and unity for the ``Add momenta'' action. In the language of Feynman diagrams, such a prescription can be interpreted as follows. The weight of each Feynman diagram is proportional to the kinematical factor
\begin{eqnarray}
\label{diag_weight}
\int d^D q_1 \ldots d^D q_{M_I}
\, \prod\limits_{i=1}^{M_I} \frac{1}{q_i^2 + m_0^2}
\, \prod\limits_{j=1}^{M_D} \frac{1}{Q_j^2 + m_0^2}
\end{eqnarray}
where $q_i$, $i = 1 \ldots M_I$ are independent momenta circulating in loops and $Q_j$, $j = 1 \ldots M_D$ can be expressed as some linear combinations of $q_i$ and the momenta of the external legs. In Algorithm \ref{alg:rp_phi4} we in fact perform Monte-Carlo integration over the independent momenta $q_i$, generating them randomly and independently at the ``Add momenta'' action with the probability distribution proportional to $1/\lr{q_i^2 + m_0^2}$. Our isotropic ultraviolet cutoff scheme thus consists in limiting the integrations over independent momenta $q_i$ to $|q_i| < \Lambda$. The factor $\chi$ in Algorithm \ref{alg:rp_phi4} is then proportional to the product of propagators involving the dependent momenta $Q_j$. The weight (\ref{diag_weight}) is obtained by averaging $\chi$ over random $q_i$.

 An alternative is to perform a Metropolis-like integration, with the probability of the ``Create vertex'' action being proportional to $1/\lr{p^2 + m_0^2}$ and with $\chi$ being identically equal to one. However, we have found that such an alternative prescription increases the rate of ``Restart'' actions quite significantly, and results in a less optimal performance of the algorithm. Therefore, we consider only the first choice, implemented in Algorithm \ref{alg:rp_phi4}.

 Let us now consider the coefficients $c_{n, m}$ more closely. As it was already shown, $c_{n, m}$ should grow as $\lr{n/2 + m}!$ at large $n$, $m$. On the other hand, if $c_{n, m}$ grow too fast, higher-order diagrams will be strongly suppressed. Let us assume that $c_{n, m}$ is proportional to $\Gamma\lr{n/2 + m + \alpha}$ times some functions which grow exponentially with $m$ and $n$. By minimizing the rate of ``Restart'' actions for all $n$ at a fixed $m$, we find the optimal value $\alpha = 1/2$. Therefore, we choose the following form of the coefficients $c_{n, m}$:
\begin{eqnarray}
\label{coeff_def}
c_{n, m} = \Gamma\lr{n/2 + m + 1/2} \, x^{-\lr{n-2}} \, y^{-m}  ,
\end{eqnarray}
with some $x$ and $y$. The sum of the probabilities of the ``Add momenta'' and ``Create vertex'' actions for such a choice of $c_{m, n}$ is:
\begin{eqnarray}
\label{coeff_ptot}
p_A + p_V = \frac{2 \, \lr{2 \pi}^D \, \Sigma_0 x^2 \lr{n+1}}{n + 2 m + 1} + \frac{y}{\lr{2 \pi}^{2 D} \, m_0^2 \, x^2}  .
\end{eqnarray}
The maximal value of the total probability is reached for $m = 0$. Minimization with respect to $x$ then shows that the total probability (\ref{coeff_ptot}) does not exceed one only if
\begin{eqnarray}
\label{y_bound}
y < \bar{y} = \frac{\lr{2 \pi}^D \, m_0^2}{8 \, \Sigma_0}
\end{eqnarray}
For $y = \bar{y}$, the total probability (\ref{coeff_ptot}) is equal to one if $x^2 = \bar{x}^2 = \frac{1}{4 \lr{2 \pi}^D \Sigma_0}$. If we set $y = \bar{y}$ and $x = \bar{x}$, as we actually did in our simulations, the probabilities of both ``Add momenta'' and ``Create vertex'' actions do not depend on the bare mass $m_0$ or on the space-time dimensionality $D$. Such a choice also minimizes the rate of ``Restart'' actions $w_R$.

 Hence, the autocorrelation time of the Markov process specified by Algorithm \ref{alg:rp_phi4} also does not depend on the physical parameters of the theory (\ref{phi4_action}). In practice, it does not exceed several iterations. We see that our algorithm does not suffer from the critical slowing-down in the sense of the usual Monte-Carlo algorithms, where autocorrelation time typically strongly increases close to the continuum limit. However, this absence of critical slowing down is compensated by the increase of the computational time which is required to obtain sufficient statistics in the low-momentum region. Indeed, the volume of this region decreases as $\Lambda_{IR}^{\lr{n-1} D}$ as the infrared cutoff $\Lambda_{IR}$ goes to zero, and hence the probability for the momenta $p_1 \ldots p_n$ to get within this region also quickly decreases.

 It is also interesting to note that the inclusive probabilities $\sum\limits_{m} G_m\lr{p_1, \ldots, p_n}$ to obtain a diagram with $n$ legs, irrespectively of the order $m$, are proportional to the Borel-Leroy transforms of the field correlators $G\lr{p_1, \ldots, p_n}$. This means that the factors $\lr{-\lambda_0}^m$ in the perturbative expansion of the correlators are replaced by $y^m / \Gamma\lr{n/2 + m + 1/2}$. Thus one can say that formally Algorithm \ref{alg:rp_phi4} stochastically estimates the Borel-Leroy transform of field correlators for $y$ in the range $0 \le y \le \bar{y}$. However, in order to recover physical results, one should integrate the Borel-Leroy transform in the range $-\infty \le y \le 0$, therefore it is difficult to use this statement for practical calculations. A much more convenient method is to analyze the dependence of the coefficients $G_m\lr{p_1, \ldots, p_n}$ on $m$, as described in Section \ref{sec:series_resummation}.

 For numerical experiments, we have implemented Algorithm \ref{alg:rp_abstract_desc} as a C program. Source code of this program is available at \cite{MyWebPage:2011}. We have used the \texttt{ranlux} random number generator at luxury level 2 \cite{Luscher:93:1}. All numerical data to which we refer below were obtained using this code. We have set $y = \bar{y}$, $x = \bar{x}$. Numerically we found that with such choice of parameters the average rate of ``Restart'' actions is $w_R = 0.282 \pm 0.001$.

\begin{figure}
  \includegraphics[width=6.0cm, angle=-90]{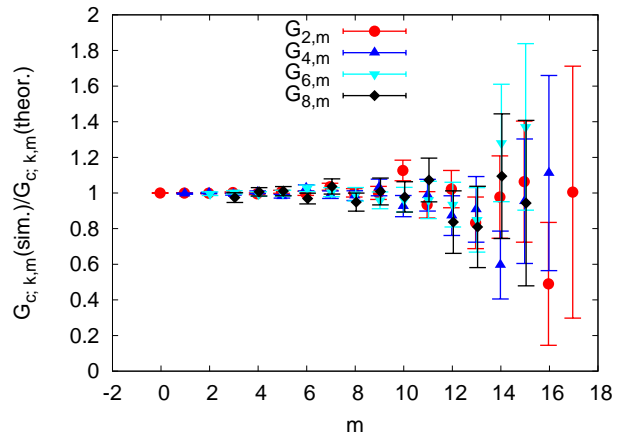}\\
  \caption{Ratios of the expansion coefficients $G_{c; n,m}$ of the connected $n$-point functions for the zero-dimensional $\phi^4$ theory computed using Algorithm \ref{alg:rp_phi4} to their exact values.}
  \label{fig:d0series_ratio}
\end{figure}

 In order to test the performance of Algorithm \ref{alg:rp_phi4}, we first consider zero-dimensional scalar field theory with the action $S\lr{\phi} = \phi^2/2 + \lambda/4 \, \phi^4$, for which the expansion coefficients in (\ref{correlators_prob_redef}) can be easily obtained in an analytic form. On Fig. \ref{fig:d0series_ratio} we plot the ratios of the expansion coefficients $G_{c; n,m}$ of the connected $n$-point functions to their exact values. These results were calculated with $5~\cdot~10^6$ iterations of our Algorithm \ref{alg:rp_phi4}. One can see that up to approximately $m=10$ numerical results differ from exact values by less than $10\%$, while at larger $m$ statistical errors become quite significant.

\begin{figure}[ht]
  \includegraphics[width=6.0cm, angle=-90]{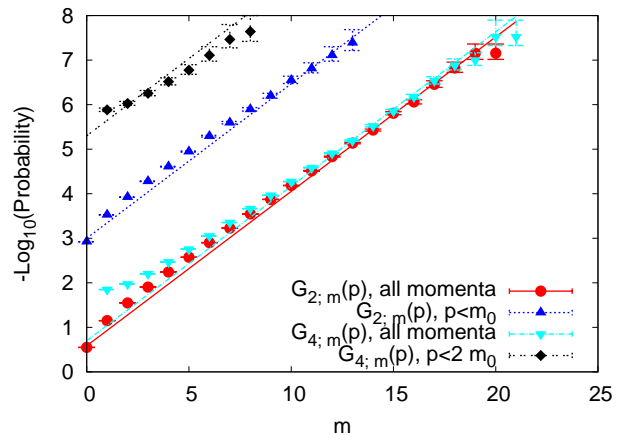}\\
  \caption{Probabilities of encountering the 2nd and 4th order connected diagrams in a random process specified by Algorithm \ref{alg:rp_phi4}, for all possible momenta and for momenta below some infrared cutoff. Solid lines are the fits of the form $A \alpha^{-m}$ with $\alpha = 0.45$}
  \label{fig:order_dist}
\end{figure}

 Next, on Fig. \ref{fig:order_dist} we plot the probabilities of encountering the two- and four-point diagrams of order $m$ per one iteration as a function of $m$ for the four-dimensional theory. The label ``all momenta'' means that the diagrams were added to statistics independently of the momenta of their external legs. The label $p < \Lambda_{IR}$, where $\Lambda_{IR} = m_0$ for the two-point diagrams and $\Lambda_{IR} = 2 m_0$ for the four-point diagrams, means that the diagrams were additionally weighted by the factor $\expa{- \sum \limits_{A=1}^{n} \frac{p_A^2}{2 \Lambda_{IR}^2} }$ (see also Section \ref{sec:series_resummation}). The data was obtained for $10^8$ iterations of Algorithm \ref{alg:rp_phi4} for $D = 4$ and $m_0 = 0.15$. We see that when the diagrams are added to statistics irrespectively of the momenta of external legs, at large order $m$ the probabilities of encountering both the two- and four-point diagrams of order $m$ are almost equal and decay as $\alpha^{-m}$, where $\alpha = 0.45 \pm 0.05$. This value is very close to the probability $p_V = 1/2$ of ``Create vertex'' action in Algorithm \ref{alg:rp_phi4}, since the probability of ``Add momenta'' action is suppressed at large $m$. Imposing the infrared cutoff results in a strong kinematical suppression of the probabilities, but they still decay as $\alpha^{-m}$ at large $m$. The fits of the form $A \, \alpha^{-m}$ are plotted on Fig. \ref{fig:order_dist} with solid lines.

\section{Resummation and infrared limit}
\label{sec:series_resummation}

 Algorithm \ref{alg:rp_phi4} stochastically estimates the coefficients of perturbative expansion of the correlators (\ref{phi4_correlators_def}), reweighted by the factors $c_{n, m}$. In order to measure some physical observable one should be able to re-sum somehow the factorially divergent perturbative series (\ref{correlators_prob_redef}). In addition, the zero-momentum limit of the correlators (\ref{phi4_correlators_def}) should be taken to measure the renormalized parameters of the theory from (\ref{renorm_mass_def}) and (\ref{renorm_coupling_def}).

\begin{figure*}[ht]
  \includegraphics[width=6.0cm, angle=-90]{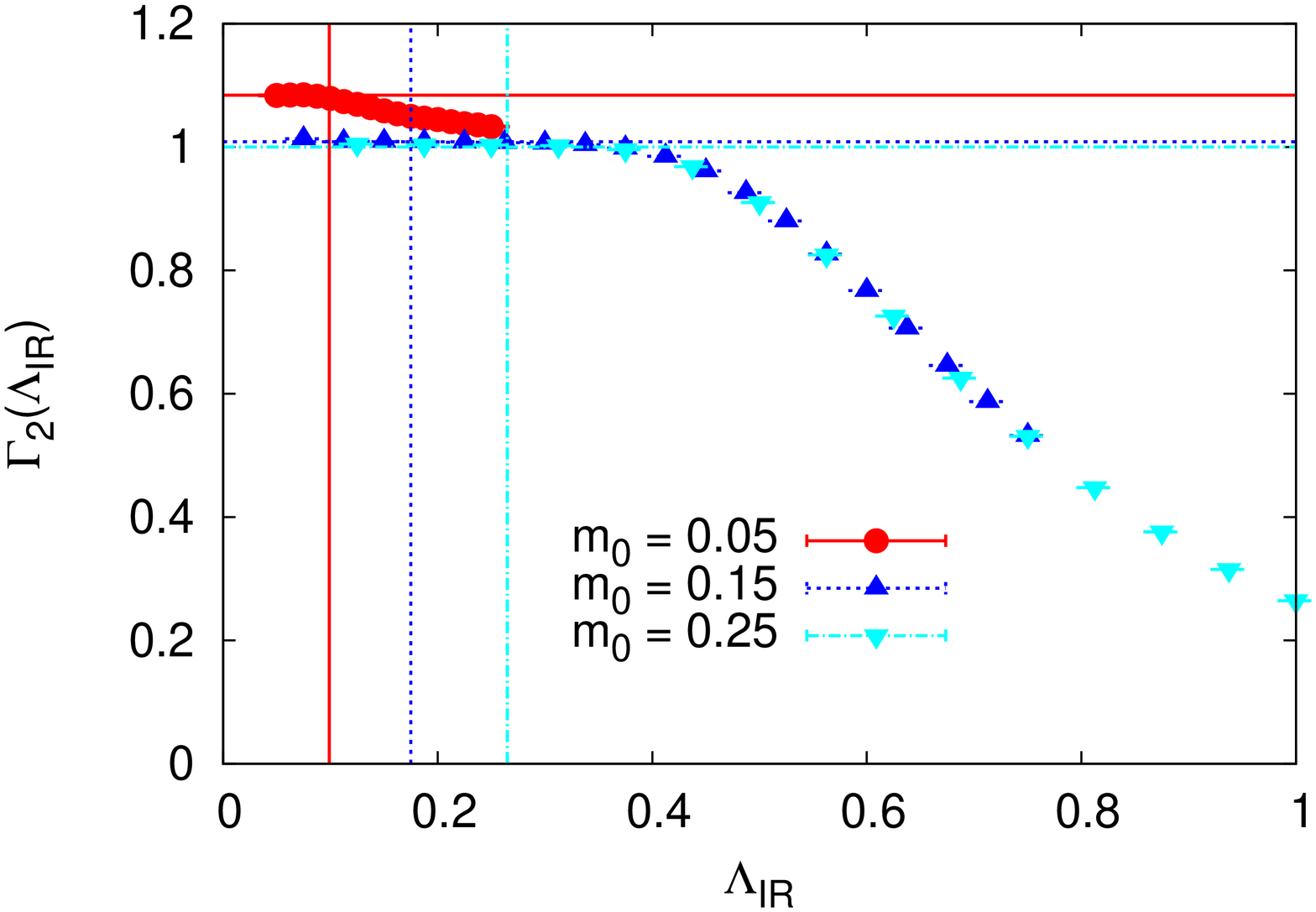}\includegraphics[width=6.0cm, angle=-90]{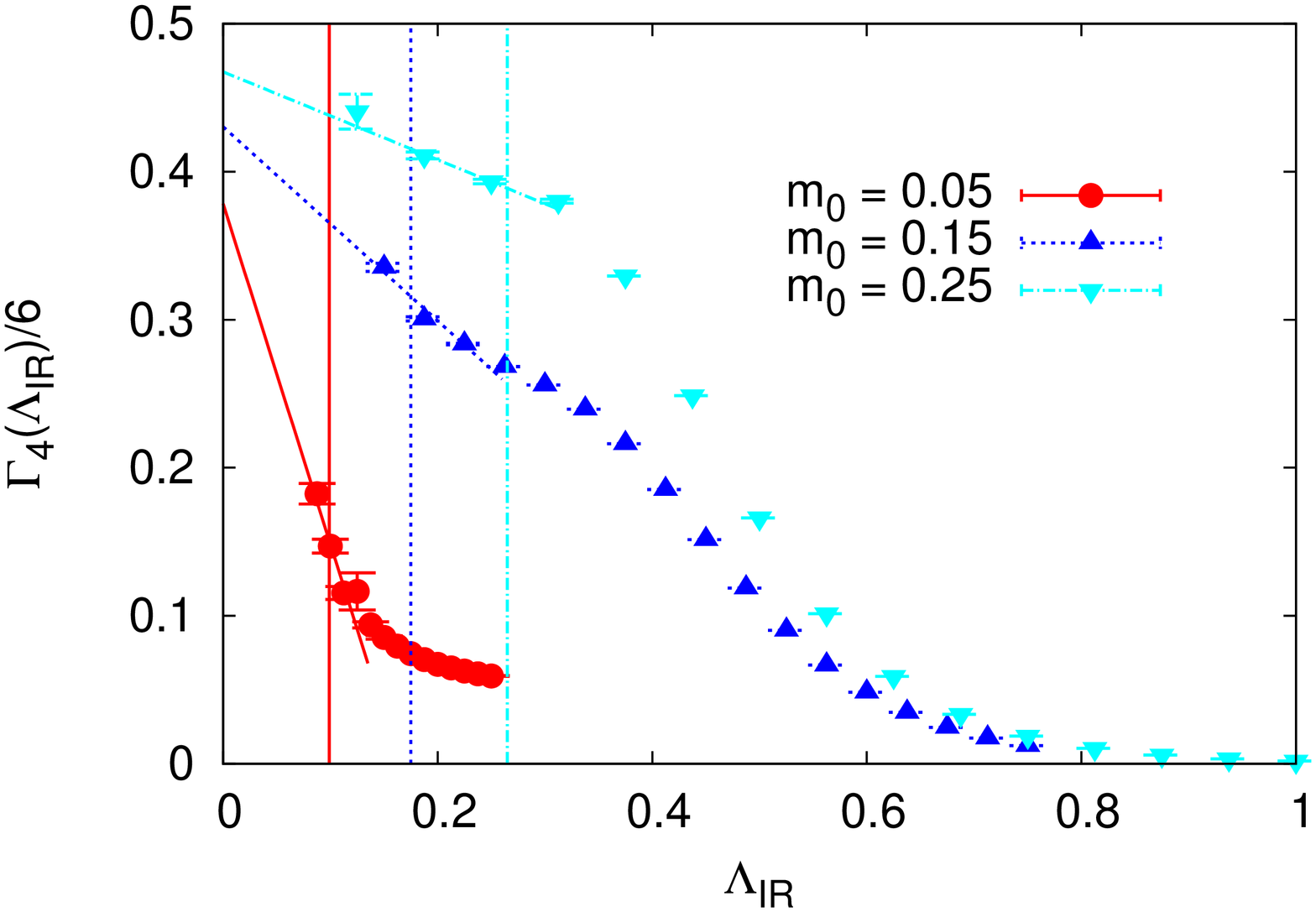}\\
  \includegraphics[width=6.0cm, angle=-90]{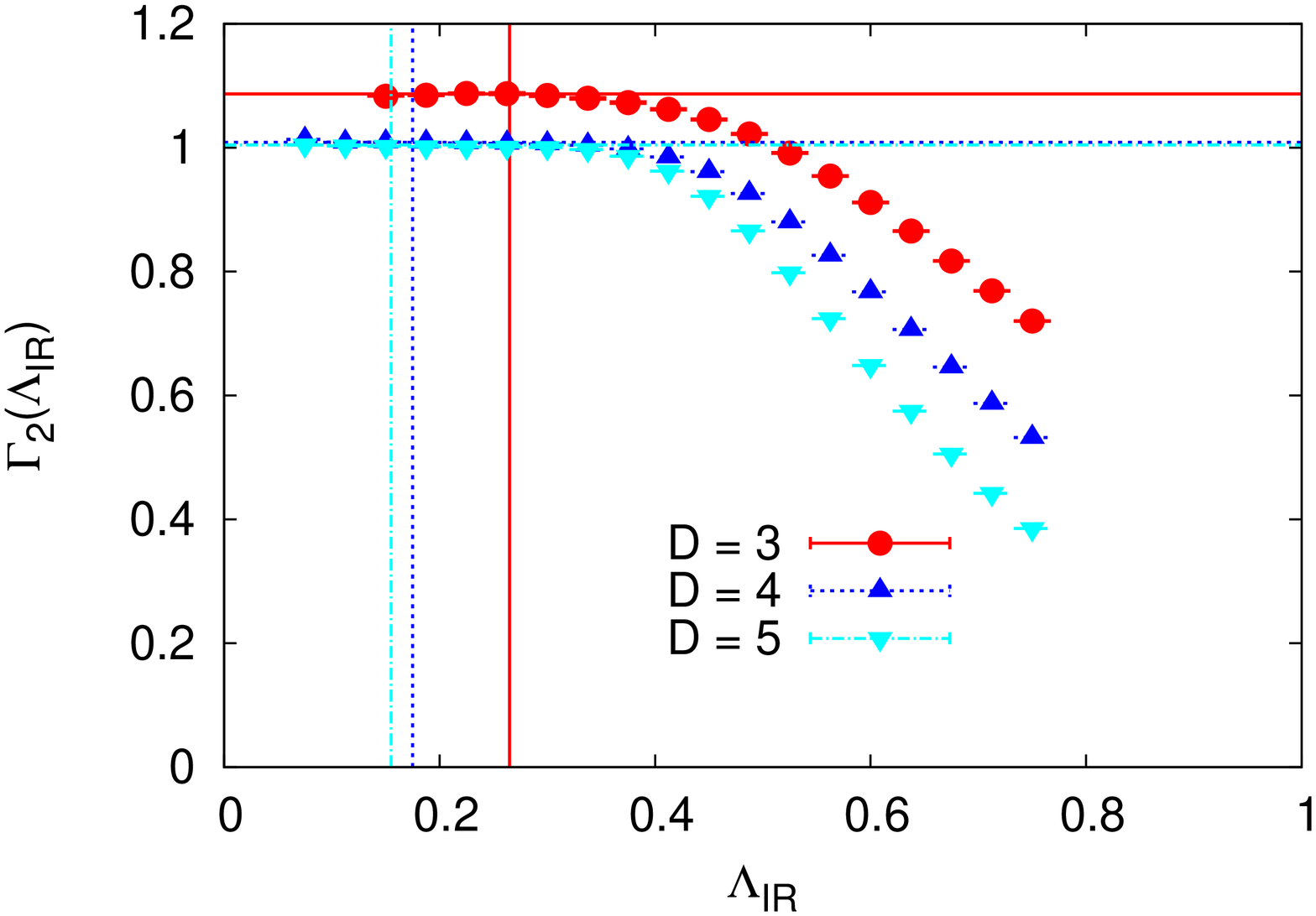}\includegraphics[width=6.0cm, angle=-90]{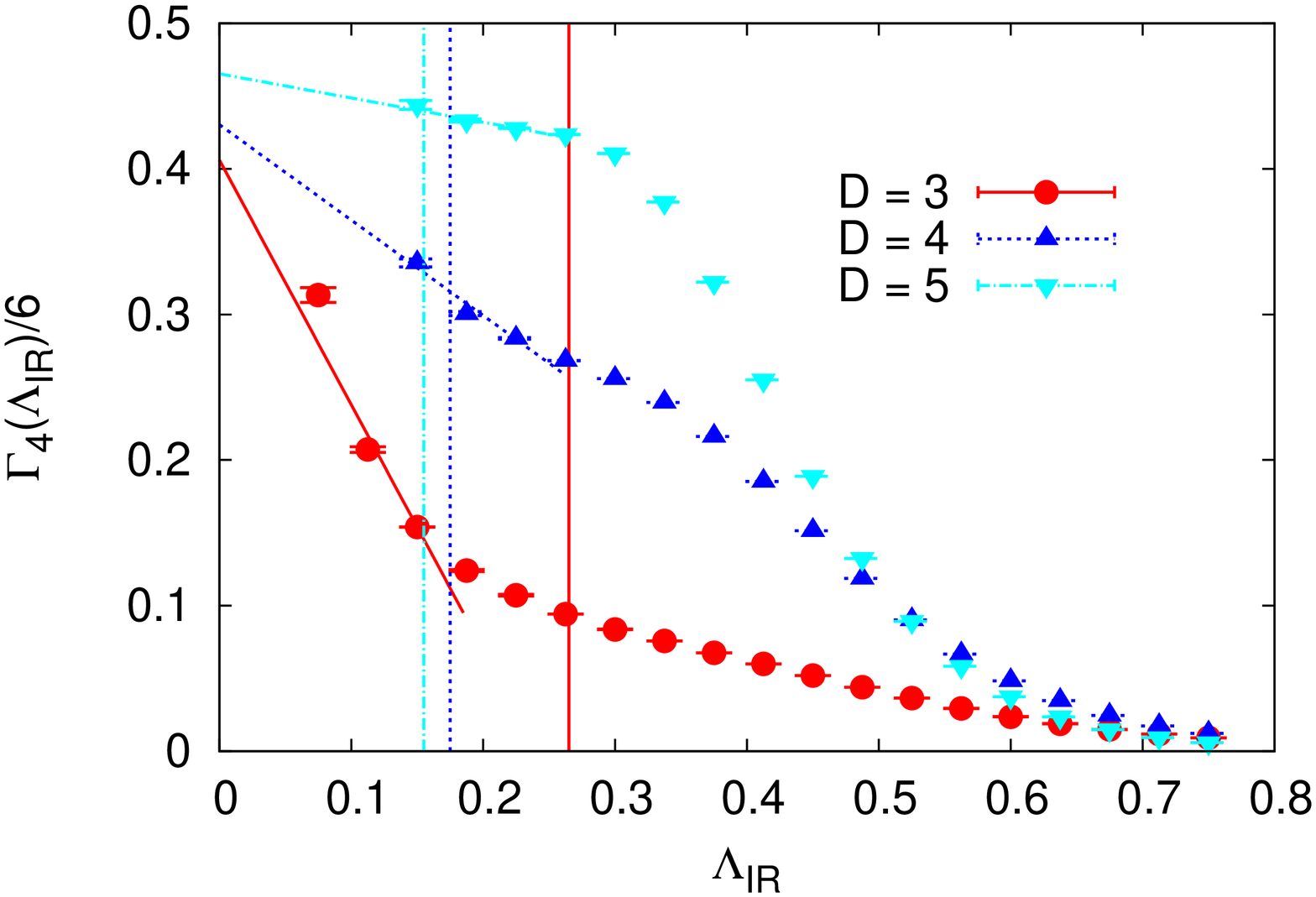}\\
  \caption{Regularized zero-momentum limits $\Gamma_2\lr{\Lambda_{IR}}$ (on the left) and $\Gamma_4\lr{\Lambda_{IR}}$ (on the right) of the two-point and four-point connected truncated correlators as functions of the infrared cutoff $\Lambda_{IR}$ for $\lambda_0 = 0.5$ (for $D = 3$, we use $m_0 = 0.15$ and $\lambda_0 = 0.2$). Above: at different bare masses at $D = 4$, below: at $m_0 = 0.15$ in different space-time dimensions. Vertical solid lines denote the scale $\Lambda_{IR} = m_R$ and the slanting and horizontal lines are linear functions which extrapolate $\Gamma_2\lr{\Lambda_{IR}}$ and $\Gamma_4\lr{\Lambda_{IR}}$ to $\Lambda_{IR} = 0$.}
  \label{fig:resummed}
\end{figure*}

  Let us first discuss the zero-momentum limit of the field correlators. We will first take the zero-momentum limit of each expansion coefficient $G_m\lr{p_1, \ldots, p_n}$ in (\ref{correlators_prob_redef}), and then consider the resummation of the series (\ref{correlators_prob_redef}). As discussed above, Algorithm \ref{alg:rp_phi4} produces sequences of momenta of the external legs of Feynman diagrams, and $G_m\lr{0, \ldots, 0}$ is proportional to the probability of all the momenta in the sequence being zero. To calculate this probability numerically, one should measure the probability for all the momenta $p_1, \ldots, p_n$ to belong to some small region near the point $p_1 = 0, \ldots, p_n = 0$, and then extrapolate the ratio of this probability to the volume of the region to the zero size of this region. Here we use soft infrared cutoff, and define for any function $F\lr{p_1, \ldots, p_n}$ of $n$ momenta
\begin{eqnarray}
\label{ir_cutoff_def}
F_n\lr{\Lambda_{IR}} = \int d^D p_1 \ldots d^D p_n
 \nonumber \\ 
\delta_{IR}\lr{p_1, \ldots, p_n; \Lambda_{IR}} \, F\lr{p_1, \ldots, p_n}  ,
\end{eqnarray}
where
\begin{eqnarray}
\label{ir_cutoff_factor}
\delta_{IR}\lr{p_1, \ldots, p_n; \Lambda_{IR}}
 =  \nonumber \\ = 
\frac{n^{D/2}}{\lr{\sqrt{2 \pi \Lambda_{IR}^2} \, }^{\lr{n-1} \, D}} \,
\expa{- \sum \limits_{A = 1}^{n} \frac{p_A^2}{2 \Lambda_{IR}^2}}
\end{eqnarray}
and we also assume that $F\lr{p_1, \ldots, p_n}$ contains the factor $\delta\lr{\sum\limits_{A=1}^{n} p_A}$. This factor should be omitted when taking the zero-momentum limit, as in the definitions of the renormalized parameters (\ref{renorm_mass_def}) and (\ref{renorm_coupling_def}). The zero-momentum limit of $F\lr{p_1, \ldots, p_n}$ corresponds then to the limit $F_n\lr{\Lambda_{IR} \rightarrow 0}$.

  In order to measure the renormalized parameters $m_R$, $\lambda_R$ and $Z_R$ from the behavior of correlators (\ref{renorm_mass_def}) and (\ref{renorm_coupling_def}), we consider the regularized zero-momentum limit (\ref{ir_cutoff_def}) $\Gamma_2\lr{\Lambda_{IR}}$ of the quantity $\Gamma\lr{p_1, p_2} = \lr{m_R^2 + p_1^2} \, G\lr{p_1, p_2}$ and tune the renormalized mass $m_R^2$ so that the deviation of $\Gamma_2\lr{\Lambda_{IR}}$ from a constant value is minimized for $\Lambda_{IR} < m_R$. This constant is by definition the wave function renormalization constant $Z_R$.

 The functions $\Gamma_2\lr{\Lambda_{IR}}$ are plotted for different bare masses and for different space-time dimensions on Fig. \ref{fig:resummed} on the left. Horizontal solid lines are the values of $Z_R$, and vertical solid lines correspond to $\Lambda_{IR} = m_R$. One can see that the limit $\Lambda_{IR} \rightarrow 0$ is indeed well-defined in this case, for all the values of the bare mass $m_0$ and for all space-time dimensions.

 In order to calculate $\Gamma_2\lr{\Lambda_{IR}}$, we have to calculate first its expansion coefficients $\Gamma_{2, m}\lr{\Lambda_{IR}}$, which are defined similarly to (\ref{correlators_prob_redef}). Taking into account the definition (\ref{ir_cutoff_def}), we measure these coefficients by summing the quantities
\begin{eqnarray}
\label{g2_average_def}
I_{2}\lr{m, l, \Lambda_{IR}} = \delta_{n,2} \delta_{IR}\lr{p_1, p_2; \Lambda_{IR}} \, \lr{p_1^2}^l ,
\end{eqnarray}
where $l = 0, 1$, separately for different diagram orders $m$ over sufficiently large number of iterations of Algorithm \ref{alg:rp_phi4}. Then
\begin{eqnarray}
\label{g2_average_final}
\Gamma_{2, m}\lr{\Lambda_{IR}} = \frac{\Sigma_0}{c_{2, 0} \, w_R} \,
 \nonumber \\ 
\lr{ \overline{I_{2}\lr{m, 1, \Lambda_{IR}}} + m_R^2 \, \overline{I_{2}\lr{m, 0, \Lambda_{IR}}}}  ,
\end{eqnarray}
where the first factor arises due to normalization of the source term in (\ref{phi4_sd_disconnected_n2}) and $\overline{I_{2}\lr{m, l, \Lambda_{IR}}}$ denotes averaging over a sufficiently large number of iterations of the Markov process specified by Algorithm \ref{alg:rp_phi4}, as described in (\ref{averaging_formula}).

 Similarly, in order to measure the renormalized coupling constant from (\ref{renorm_coupling_def}), we should measure the one-particle irreducible four-point correlator $\Gamma\lr{p_1, p_2, p_3, p_4}$ at zero external momenta. At small external momenta we use (\ref{renorm_mass_def}) and (\ref{renorm_coupling_def}) and write
\begin{eqnarray}
\label{four_point_func_approx}
\Gamma\lr{p_1, p_2, p_3, p_4} \approx
 \nonumber \\ 
\prod \limits_{A=1}^{4} Z_R^{-1} \, \lr{p_A^2 + m_R^2} G_c'\lr{p_1, p_2, p_3, p_4}  .
\end{eqnarray}
Now one can take the regularized zero-momentum limit of the expansion coefficients of (\ref{four_point_func_approx}) by summing the quantities
\begin{eqnarray}
\label{g4_average_def}
I_{4}\lr{m, l, \Lambda_{IR}}
 = \nonumber \\ = 
\delta_{n, 4} \, \delta_{IR}\lr{p_1, p_2, p_3, p_4; \Lambda_{IR}} \, \xi_l\lr{p_1, p_2, p_3, p_4}
\end{eqnarray}
separately for connected diagrams of different orders $m$ over sufficiently large number of iterations of Algorithm \ref{alg:rp_phi4}, as in (\ref{averaging_formula}). Here we have defined the kinematical factors $\xi_l\lr{p_1, p_2, p_3, p_4}$, $l = 0, \ldots, 4$ as follows:
\begin{eqnarray}
\label{g4_momenta_combs}
\xi_0\lr{p_1, p_2, p_3, p_4} = p_1^2 \, p_2^2 \, p_3^2 \, p_4^2
\nonumber \\
\xi_1\lr{p_1, p_2, p_3, p_4} = p_2^2 \, p_3^2 \, p_4^2 + p_1^2 \, p_3^2 \, p_4^2
 +  \nonumber \\ + 
p_1^2 \, p_2^2 \, p_4^2 + p_1^2 \, p_2^2 \, p_3^2
\nonumber \\
\xi_2\lr{p_1, p_2, p_3, p_4} = p_1^2 \, p_2^2 + p_1^2 \, p_3^2 + p_1^2 \, p_4^2
 +  \nonumber \\ + 
p_2^2 \, p_3^2 + p_2^2 \, p_4^2 + p_3^2 \, p_4^2
\nonumber \\
\xi_3\lr{p_1, p_2, p_3, p_4} = p_1^2 + p_2^2 + p_3^2 + p_4^2
\nonumber \\
\xi_4\lr{p_1, p_2, p_3, p_4} = 1
\end{eqnarray}
Then the expansion coefficients of the regularized zero-momentum limit of the truncated connected four-point correlator (\ref{four_point_func_approx}) are given by
\begin{eqnarray}
\label{g4_average_final}
\Gamma_{4, m}\lr{\Lambda_{IR}} = \frac{\Sigma_0}{c_{2,0} \, w_R} \, \sum \limits_{l = 0}^{4} \, m_R^{2 l} \, \overline{I_{4}\lr{m, l, \Lambda_{IR}}}
\end{eqnarray}

\begin{figure}[ht]
  \includegraphics[width=6.0cm, angle=-90]{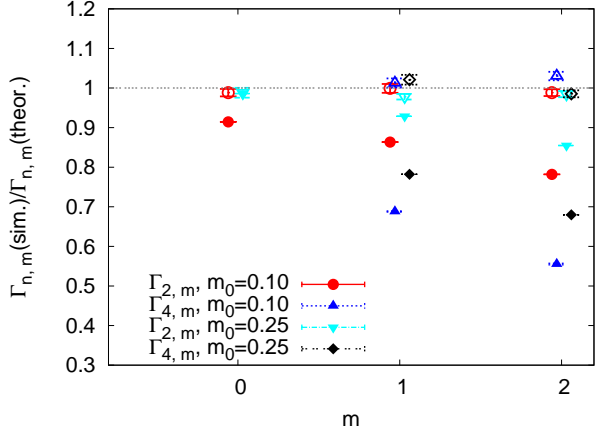}\\
  \caption{Ratios of the measured expansion coefficients $\Gamma_{n, m}\lr{\Lambda_{IR}}$ to their analytical values for different bare masses $m_0$ in a four-dimensional theory. For the two- and four-point functions we take $\Lambda_{IR} = m_0/4$ and $\Lambda_{IR} = m_0$, respectively. For the data points shown with full symbols, external momenta were put to zero in analytical expressions. For data points shown with empty symbols, external momenta were integrated over with the weight $\delta_{IR}\lr{p_1, \ldots, p_n; \Lambda_{IR}}$.}
  \label{fig:low_order_contrib}
\end{figure}

 The resummed function $\Gamma_4\lr{\Lambda_{IR}}$ is plotted for different bare masses and for different space-time dimensions on Fig. \ref{fig:resummed} on the right (to make comparison with (\ref{renorm_coupling_def}) easier, we divide it by $6$). Since the low-momentum region for the four-point correlator is much stronger suppressed kinematically then for the two-point correlator, the limit $\Lambda_{IR} \rightarrow 0$ is now not so well-defined, especially at small bare mass $m_0$. In order to extrapolate $\Gamma_4\lr{\Lambda_{IR}}$ to zero, we fit several (from 3 to 5) data points with smallest $\Lambda_{IR}$ by a linear function, and assume that the intercept of this linear function is the required limit $\Gamma_4\lr{\Lambda_{IR} \rightarrow 0}$. Slanting solid lines on the plots on the right of Fig. \ref{fig:resummed} are these linear fits, and vertical solid lines again correspond to $\Lambda_{IR} = m_R$. Clearly, such extrapolation procedure introduces quite large systematic errors into our measurements of the renormalized coupling constant (\ref{renorm_coupling_def}). However, as we shall see below, even with such a crude extrapolation we get reasonable results for the scale dependence of $\lambda_R$.

 In order to illustrate how the soft infrared cutoff (\ref{ir_cutoff_def}) reproduces the zero-momentum limit, on Fig. \ref{fig:low_order_contrib} we also show the ratios of the expansion coefficients of the regularized zero-momentum limit of the connected $n$-point functions $\Gamma_{2, m}\lr{\Lambda_{IR}}$ and $\Gamma_{4, m}\lr{\Lambda_{IR}}$ to the results of analytical calculations up to $m = 2$. Namely, we calculate analytically perturbative contributions to the two-point function $G\lr{p}$ and the connected four-point function $G_c\lr{p_1, p_2, p_3, p_4}$ up to $m = 2$ and multiply the external legs by $\lr{m_R^2 + p_A^2}$, where $m_R^2$ is calculated from our numerical data, as described below. Then we either set all the external momenta to zero (the corresponding points on the plot are shown with full symbols) or integrate over them with the weight $\delta_{IR}\lr{p_1, \ldots, p_n; \Lambda_{IR}}$ (these points are shown with empty symbols). We set $\Lambda_{IR} = m_0/4$ for the two-point function and $\Lambda_{IR} = m_0$ for the four-point function. First we note that when both the analytical and the numerical results are weighted with the factor $\delta_{IR}\lr{p_1, \ldots, p_n; \Lambda_{IR}}$, the ratio is very close to one, which suggests that the loop integrals in (\ref{diag_weight}) are indeed accurately reproduced by our Algorithm \ref{alg:rp_phi4} for sufficiently large external momenta. On the other hand, the exact zero-momentum limit differs from the regularized result (\ref{ir_cutoff_def}). As discussed above, for the four-point function we have to take quite large IR cutoff in order to gain sufficient statistics, and the zero-momentum limit is then accessed using linear extrapolation.

 To get a deeper insight into possible problems with the regularized zero-momentum limit (\ref{ir_cutoff_def}) of the two-point function, let us consider 1-particle reducible diagrams with two external legs and with $m$ insertions of 1-particle irreducible self-energy diagrams. The contribution of such diagrams is proportional to the kinematical factor
\begin{eqnarray}
\label{nested_diagrams}
\lr{\frac{1}{p^2 + m_0^2}}^{m} \, \Sigma\lr{p}^{m-1}  ,
\end{eqnarray}
where $\Sigma\lr{p}$ is the self-energy. If we neglect the dependence of $\Sigma\lr{p}$ on $p$ (which is indeed weak for the lowest-order perturbative contributions) and fix the infrared cutoff in (\ref{ir_cutoff_def}) to some finite $\Lambda_{IR}$, at sufficiently large $m > m_0^2/\Lambda_{IR}^2$, a simple estimate shows that the regularized infrared limit (\ref{ir_cutoff_def}) of (\ref{nested_diagrams}) differs from the exact zero-momentum limit by a factor $\sim m^{-d/2}$. In our simulations the minimal value of the infrared cutoff is $\Lambda_{IR} = m_0/4$. According to the above arguments, this allows to re-sum the diagrams with up to $\sim 16$ insertions of self-energy diagrams. However, even for smaller $m$ we expect that the contributions of the rapidly changing kinematical factors of the form similar to (\ref{nested_diagrams}) are the main source of systematic errors in our simulations. According to Fig. \ref{fig:low_order_contrib}, they can be as large as $\sim 20 \%$ for $m_0 = 0.1$ and $m = 2$. Presumably, these errors can be significantly reduced by implementing some resummation procedure for one-particle reducible diagrams.

 After having discussed the zero-momentum limit, let us describe the integral Borel-Leroy transformation which we use for the resummation of the series (\ref{correlators_prob_redef}). For generality, in the following we will denote by $G_n$ any $n$-point correlator, probably multiplied by some function of momenta or integrated over all the momenta with some weight, as in (\ref{g2_average_final}) and (\ref{g4_average_final}). Correspondingly, by $G_{n, m}$ we denote the coefficients of expansion of $G_n$ in powers of the bare coupling constant $\lambda_0$, reweighted by the factors $c_{n, m}$, as in (\ref{correlators_prob_redef}).

 Inserting the explicit form of the factors $c_{n, m}$ from (\ref{coeff_def}) into the series (\ref{correlators_prob_redef}) and omitting the dependence on momenta, as discussed above, we obtain:
\begin{eqnarray}
\label{resummation1}
G_n = x^{-n+2} \sum \limits_{m=0}^{+\infty}
\Gamma\lr{n/2 + m + 1/2} \, \lr{-\frac{\lambda_0}{y}}^{m} G_{n, m}
\end{eqnarray}
Using the integral representation of the gamma-function $\Gamma\lr{x} = \int\limits_{0}^{+\infty} dt \, e^{-t} \, t^{x-1}$ and changing the integration variable to $z = \frac{\lambda_0 \, t}{y}$, we get:
\begin{eqnarray}
\label{resummation}
G_n = x^{-n+2} \, \lr{\frac{y}{\lambda_0}}^{\frac{n+1}{2}} \,
\int \limits_{0}^{+\infty} dz \, \expa{-\frac{y z}{\lambda_0}} \,
\nonumber \\ 
z^{\frac{n-1}{2}} \, \lr{ \sum \limits_{m=0}^{+\infty} \, \lr{-z}^m \, G_{n, m}}  .
\end{eqnarray}

 In real simulations, the coefficients $G_{n, m}$ are known only up to some finite maximal order. A standard resummation strategy in this case is to construct the Pade approximant of the function $G_{n}\lr{z} = \sum \limits_{m=0}^{+\infty} \, \lr{-z}^m \, G_{n, m}$, that is, to approximate $G_n\lr{z}$ by some rational function of $z$. In this case the expansion coefficients (\ref{correlators_prob_redef}) are approximated by the sum of several exponents:
\begin{eqnarray}
\label{expansion_coeff_fits}
 G_{n, m} = \sum \limits_{k} a_k b_k^{m - m_n} ,
\end{eqnarray}
where $m_n$ is the order of the tree-level diagrams which contribute to the connected $n$-point correlator: $m_2 = 0$ and $m_4 = 1$. Inserting the expression (\ref{expansion_coeff_fits}) into the integral  (\ref{resummation}), we obtain:
\begin{eqnarray}
\label{resummation_afterfit}
G_n = \lr{-1}^{m_n} \, x^{-n+2} \, \lr{\frac{y}{\lambda_0}}^{\frac{n+1}{2}}
\nonumber \\ 
\int \limits_{0}^{+\infty} dz \, \expa{-\frac{y z}{\lambda_0}} \,
z^{\frac{n-1}{2} + m_n} \, \sum\limits_k \frac{a_k}{1 + b_k z}
\end{eqnarray}
Thus each exponent in (\ref{expansion_coeff_fits}) corresponds to a simple pole of the Borel-Leroy transform. Integration over $z$ can be now performed analytically, and we obtain for the two-point correlator:
\begin{eqnarray}
\label{resummation_final_G2}
G_2 = \sum\limits_k \sqrt{\pi} a_k r_k \, \lr{1 - \sqrt{\pi r_k}\, e^{r_k} \erfc{\sqrt{r_k}} }  ,
\end{eqnarray}
where $r_k = \frac{b_k \, y}{\lambda_0}$. For the four-point correlator, the integration yields:
\begin{eqnarray}
\label{resummation_final_G4}
G_4 = x^{-2} \, \sum\limits_k \sqrt{\pi} a_k b_k
\nonumber \\ 
\lr{\lr{r_k - 1/4}^2 + 11/16 - \sqrt{\pi r_k}\,r_k^2\,e^{r_k} \erfc{\sqrt{r_k}}}  .
\end{eqnarray}

\begin{figure}[ht]
  \includegraphics[width=6.0cm, angle=-90]{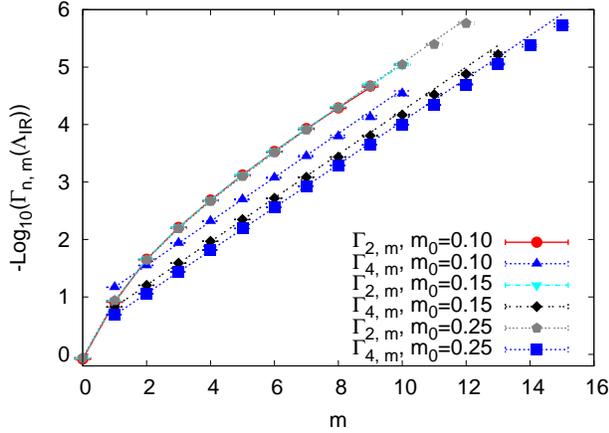}\\
  \caption{Reweighted expansion coefficients for the regularized zero-momentum limit of the two- and four-point connected correlators $\Gamma_{2, m}\lr{\Lambda_{IR} = m_0}$ (\ref{g2_average_final}) and $\Gamma_{4, m}\lr{\Lambda_{IR} = 2 m_0}$ (\ref{g4_average_final}) for different values of the bare mass $m_0$ in the four-dimensional theory. Solid lines are the fits of the data with the sum of exponential functions (\ref{expansion_coeff_fits}). Only the coefficients with relative error below $10\%$ were used for fitting and are shown on the plot.}
  \label{fig:order_contrib}
\end{figure}

 The expansion coefficients $\Gamma_{2, m}\lr{\Lambda_{IR} = m_0}$ and $\Gamma_{4, m}\lr{\Lambda_{IR} = 2 m_0}$ for the four-dimensional theory, defined by (\ref{g2_average_final}) and (\ref{g4_average_final}), are plotted on Fig. \ref{fig:order_contrib} for different values of the bare mass $m_0$. For this plot, we also omit the $m$-independent factor $x^{-\lr{n-2}}$ in (\ref{coeff_def}). The renormalized mass $m_R$ used to calculate (\ref{g2_average_final}) and (\ref{g4_average_final}) corresponds to $\lambda_0 = 0.1$. Solid lines are the fits of the form (\ref{expansion_coeff_fits}), with three exponents for the two-point correlators and with one exponent - for the four-point correlators.

\begin{figure*}[ht]
  \includegraphics[width=6cm, angle=-90]{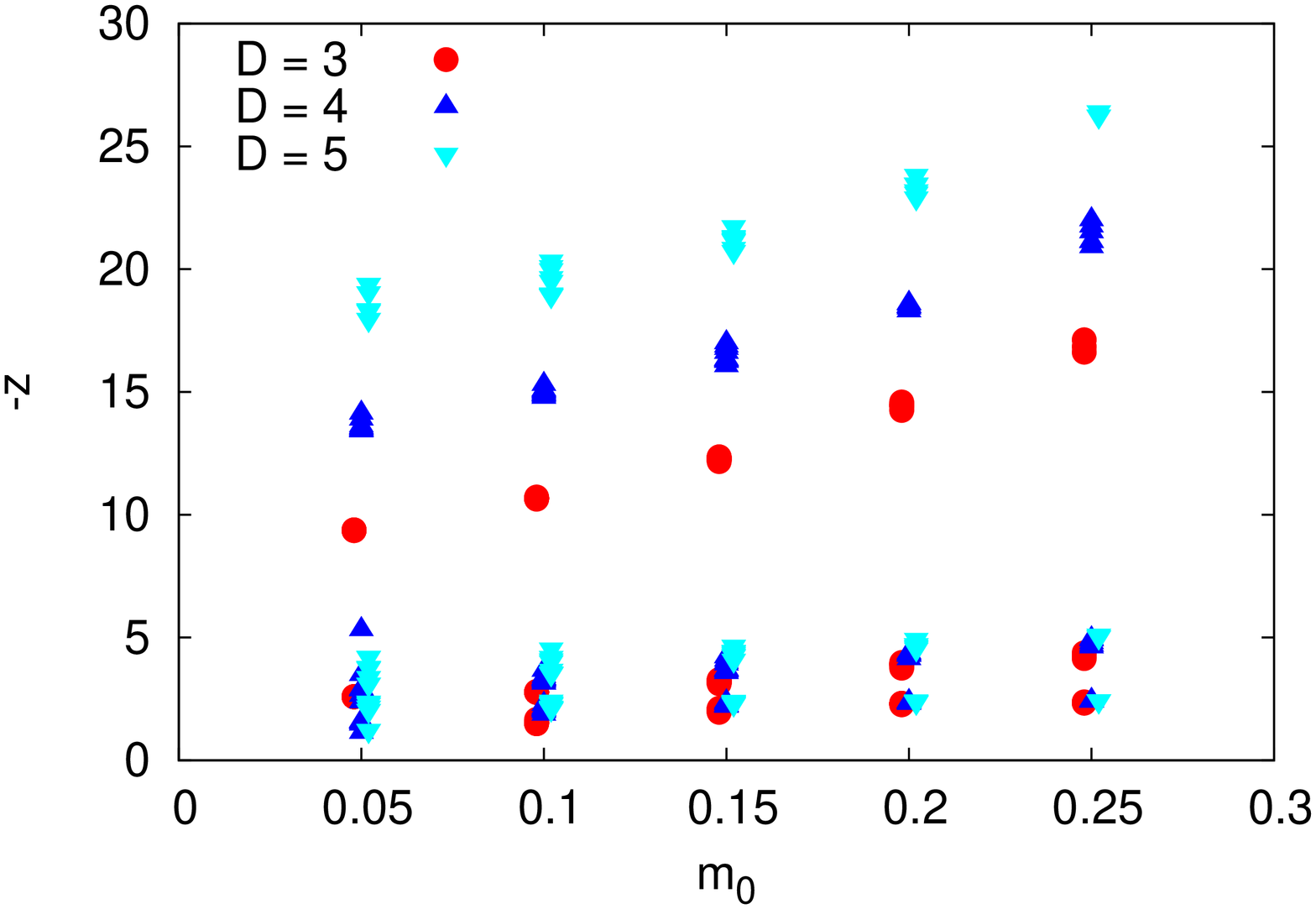}
  \includegraphics[width=6cm, angle=-90]{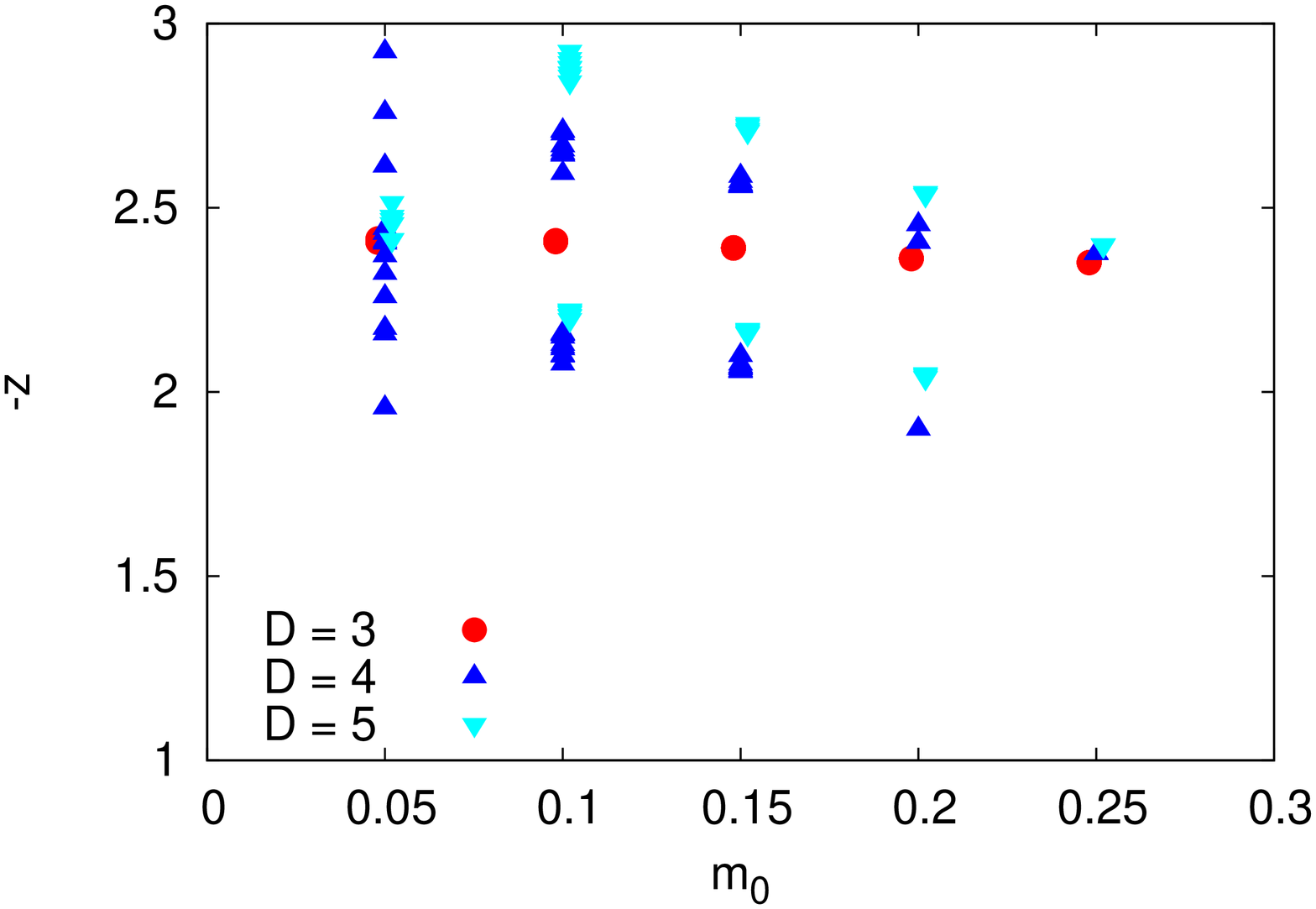}\\
  \caption{Positions of the poles of the Borel-Leroy transforms of the truncated connected two- and four-point correlators in the low-momentum region $\Gamma_2\lr{\Lambda_{IR} = m_0}$ (on the left) and $\Gamma_4\lr{\Lambda_{IR} = 3 m_0}$ (on the right) on the negative $z$ axis as a function of the bare mass $m_0$, for different space-time dimensions. For better visibility, data points for different $D$ are displaced a bit along the horizontal axis.}
  \label{fig:poles}
\end{figure*}

 Due to statistical errors in the coefficients $G_{n, m}$, the standard Pade approximant constructed on all the available data points turns out to be very unstable - small errors in the data lead to very large deviations of the approximant and to appearance of multiple spurious poles. In order to obtain more stable results, we find the optimal number of exponents and the values of $a_k$ and $b_k$ by fitting the expression (\ref{expansion_coeff_fits}) to the numerical data.

 A technical difficulty is that such fits typically result in badly conditioned minimization problems, if the number of exponents exceeds two. Fortunately, there are specific methods which work well for this particular problem. They are based on the singular value decomposition of the Hankel matrix or the so-called Page matrix \cite{DeGroen:87:1}. We found that for our data the fits based on the Hankel matrices are optimal. Let us briefly describe this fitting procedure. Define the Hankel matrix $H_{kl}$ and the shifted Hankel matrix $\bar{H}_{kl}$ as
\begin{eqnarray}
\label{hankel_matrices_def}
H_{kl} = G_{n, k + l + m_n}
\nonumber \\
\bar{H}_{kl} = G_{n, k + l + m_n + 1},
\quad  \nonumber \\ 
k, l = 0 \ldots \lfloor \lr{m_{\rm max} - m_n}/2 \rfloor - 1,
\end{eqnarray}
where $m_{\rm max}$ is the maximal order to which the expansion coefficients are known and $\lfloor \ldots \rfloor$ is the floor function. The Hankel matrix can be decomposed as $H = U \Sigma V^{T}$, where $\Sigma$ is the diagonal matrix with positive elements which are assumed to decrease from left to the right. We now form another matrix $M = \lr{\Sigma^{-1/2} U^{T} \bar{H} V \Sigma^{-1/2}}_{kl}$ of rank $N$, with $k, l = 0 \ldots N-1$, where $N$ is the required number of exponents in (\ref{expansion_coeff_fits}). The eigenvalues of $M$ are then the optimal values of the coefficients $b_k$ in (\ref{expansion_coeff_fits}) \cite{DeGroen:87:1}. When $b_k$ are known, the optimal values of $a_k$ can be easily found by a simple linear regression.

 In our simulations we have used such maximal number of exponents $N$, for which all $b_k$ in (\ref{expansion_coeff_fits}) are still real and positive, so that all the poles of the Borel-Leroy transforms of field correlators (\ref{resummation_afterfit}) lie on the real $z$ axis at $z < 0$. We have found that in this case the positions of all the poles are numerically stable. On the other hand, if we allow also for poles at $z > 0$ or for poles off the real axis, their positions turn out to be numerically very unstable. We therefore disregard them as numerical artifacts. There are also theoretical arguments \cite{ZinnJustin:81:1} that the Borel image of field correlators in bare perturbation theory should not have poles at positive real $z$.

 A disadvantage of the fitting method described above is that it does not take into account the statistical errors in the data when finding the optimal values of $b_k$ - the weights of data points do not depend on their errors. Therefore, after having found the optimal number of parameters and their values in (\ref{expansion_coeff_fits}) from the singular value decomposition of the Hankel matrix, we use this values as an initial guess for the standard minimization-based fitting algorithm. In most cases, the optimal values which take into account the errors in the data turn out to be very close to the ones that were found from the Hankel matrices, and the minimization procedure quickly converges. The resulting $\chi^2/d.o.f$ is then typically of order of unity. We also note that we have used only the coefficients with relative error below $10\%$ for fitting.

 The positions of the poles of the Pade approximants of the Borel-Leroy transforms of the functions $\Gamma_{2}\lr{\Lambda_{IR} = m_0}$ and $\Gamma_{4}\lr{\Lambda_{IR} = 3 m_0}$ on the negative $z$ axis are shown on Fig. \ref{fig:poles} as a function of the bare mass $m_0$. In order to illustrate the uncertainties in pole positions, in Fig. \ref{fig:poles} we show a scatter plot of poles for $10$ statistically independent data sets, each obtained with $10^9$ iterations of Algorithm \ref{alg:rp_phi4}. For the two-point correlator, our fitting procedure reproduces three distinct poles, and for the four-point connected correlator, for which numerical errors are more significant, from one to two distinct poles can be seen. Note that when the precision of the numerical data is not sufficient to find two distinct poles, our fitting procedure yields only one pole which is situated between the two poles which would be found if the precision of the data would be higher. This can be clearly seen for the smallest value of the bare mass $m_0 = 0.05$. At small bare mass the positions of the poles are also less stable numerically, and hence larger statistics is required to find them with good precision.

\section{Physical results}
\label{sec:phys_res}

\begin{figure}[ht]
  \includegraphics[width=6cm, angle=-90]{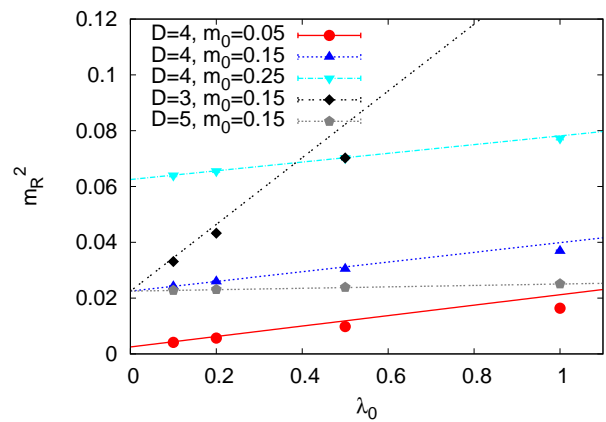}\\
  \caption{Renormalized mass $m_R$ as a function of the bare coupling constant $\lambda_0$ for different bare masses $m_0$ and in different space-time dimensions. Solid lines correspond to the one-loop contribution to $m_R$.}
  \label{fig:mr_vs_l}
\end{figure}

\begin{figure}[ht]
  \includegraphics[width=6cm, angle=-90]{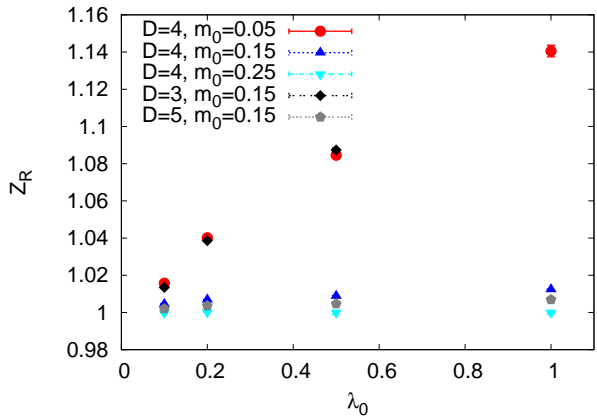}\\
  \caption{Wave function renormalization constant $Z_R$ as a function of bare coupling $\lambda_0$ for different bare masses and in different space-time dimensions.}
  \label{fig:zr_vs_l}
\end{figure}

\begin{figure}[ht]
  \includegraphics[width=6.0cm, angle=-90]{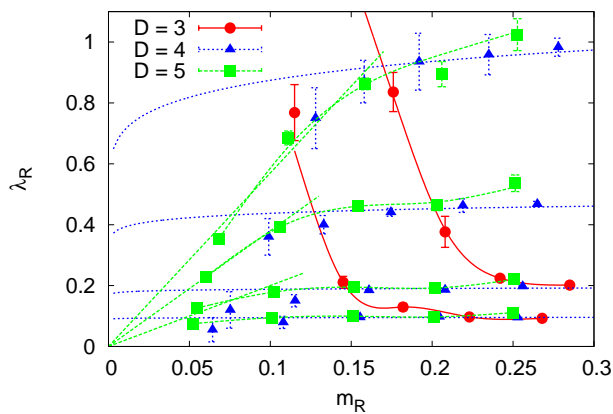}\\
  \caption{Renormalized coupling constant as a function of renormalized mass for fixed bare coupling $\lambda_0$. For $D = 4$ and $D = 5$ we take $\lambda_0 = 0.1, \, 0.2, \, 0.5, \, 1.0$ and for $D = 3$, $\lambda_0 = 0.1, 0.2$. For $D = 4$, solid lines correspond to the result of integration of one-loop $\beta$-function. For $D = 5$, straight solid lines are the linear fits of the data at small masses. For $D = 3$ and $D = 5$, interpolating splines are also shown to guide the eye.}
  \label{fig:RGflows}
\end{figure}

 In this Section we present the results of the measurements of the renormalized parameters of the theory - the renormalized mass $m_R$, the renormalized coupling constant $\lambda_R$ and the wave function renormalization constant $Z_R$ as defined in (\ref{renorm_mass_def}) and (\ref{renorm_coupling_def}). The resummation procedure and the regularized zero-momentum limit, which were used to obtain these results, are described in Section \ref{sec:series_resummation}. All the results presented on Figs. \ref{fig:order_contrib} - \ref{fig:RGflows} were obtained by averaging over $10$ independent runs of Algorithm \ref{alg:rp_phi4}, each consisting of $10^9$ iterations for $D = 3, 4$ and $4 \cdot 10^9$ iterations for $D = 5$. For $D = 5$, such increase of the number of iterations was motivated by stronger kinematical suppression of low-momentum region in higher space-time dimension. Averaging over independent runs was made in order to accurately estimate the statistical errors in the resummed correlators (\ref{resummation_final_G2}) and (\ref{resummation_final_G4}). Generation of each data set with $10^9$ iterations took several hours on a single 2 GHz CPU, which is comparable to the computer time which was required to produce similar results using the ``worm'' algorithm (several core-months for $\sim 20$ data points in different dimensions \cite{Wolff:09:1}). From Fig. \ref{fig:order_dist} one can see that with such statistics from $10$ to $15$ orders of perturbative expansion of field correlators can be analyzed in the small-momentum region (we use only coefficient with relative errors below $10\%$ for resummation, as discussed above).

 On Fig. \ref{fig:mr_vs_l} we illustrate the dependence of the renormalized mass $m_R$ on the bare coupling constant $\lambda_0$ for different space-time dimensions and for different bare masses $m_0$. In order to demonstrate that our resummation procedure yields the results which agree with the lowest orders of perturbation theory at small $\lambda_0$, on Fig. \ref{fig:mr_vs_l} we also plot the one-loop result for the renormalized mass, which indeed fits the data for small $\lambda_0$.

 Fig. \ref{fig:zr_vs_l} shows the dependence of the wave function renormalization constant $Z_R$ on the bare coupling constant. $Z_R$ is close to unity in the whole range of coupling constants and for all dimensions $D$, which agrees with the results of direct Monte-Carlo simulations \cite{Drummond:87:1}. Two-loop perturbative calculation of the self-energy shows that $Z_R < 1$, with deviation from unity not exceeding $10^{-3}$ for $\lambda_0 < 1.0$. On the other hand, our calculations show that $Z_R > 1$. Most likely this difference is due to systematic errors in our approximation of the zero-momentum limit of field correlators, which were discussed above in detail.

 Finally, on Fig. \ref{fig:RGflows} we illustrate the scale dependence of the renormalized coupling constant. Namely, we fix the bare coupling constant $\lambda_0$ (we use $\lambda_0 = 0.1, \, 0.2, \, 0.5, \, 1.0$ for $D = 4, 5$ and $\lambda_0 = 0.1, \, 0.2$ for $D = 3$) and change the bare mass $m_0$. The renormalized coupling constant $\lambda_R$ is then studied as a function of the renormalized mass $m_R$ in units of UV cutoff. This is, of course, equivalent to fixing the value of $m_R$ in physical units and changing the ultraviolet cutoff $\Lambda$. According to the renormalization group arguments \cite{Aizenman:81:1, Frohlich:82:1}, when the continuum limit of the theory is approached and the renormalized mass $m_R$ in units of UV cutoff tends to zero, in space-time dimension $D = 4$ or larger the renormalized coupling should tend to zero. Hence the continuum limit of the $\phi^4$ theory in $D \ge 4$ dimensions is a free theory of massless scalar fields.

 Our data confirms this triviality conjecture: for both $D = 4$ and $D = 5$ the renormalized coupling clearly decreases with the renormalized mass. For $D = 4$ the solid line on Fig. \ref{fig:RGflows} is the result of integration of the one-loop $\beta$-function, which implies that $\lambda_R$ should approach zero very slowly, at a logarithmic rate. Our results are consistent with such behavior within error range, although at small bare mass $\lambda_R$ seems to decrease faster than logarithmically. This systematic deviation from logarithmic scaling is probably due to large systematic errors in the measurement of $\lambda_R$ at small $m_0$, as discussed in Section \ref{sec:series_resummation}. For $D = 5$, at small masses the renormalized coupling goes to zero almost linearly, in agreement with the dimensional analysis. Linear fits of $\lambda_R$ on several data points at small renormalized mass are also shown as solid lines on Fig. \ref{fig:RGflows}.

 On the other hand, for $D = 3$ renormalization-group arguments predict that the trivial fixed point at $m_R = 0$, $\lambda_R = 0$ is unstable \cite{Frohlich:82:1}. Our data agrees with this statement: the renormalized coupling in this case quickly grows as $m_R$ goes to zero. At large renormalized mass $m_R$ the renormalized coupling tends to its bare value $\lambda_0$ for all bare masses and space-time dimensions.

\section{Discussion and conclusions}
\label{sec:conclusions}

 In this paper we have presented a novel simulation method which stochastically samples open Feynman diagrams with probability which is proportional to their weight times some re-summing combinatorial factor. In this respect, our method is similar to the ``worm'' algorithm by Prokof'ev and Svistunov \cite{Prokofev:98:1}, which is often used in the framework of Diagrammatic Monte-Carlo. The basic idea behind our approach is the stochastic perturbative solution of Schwinger-Dyson equations, which form an infinite system of linear inhomogeneous equations. Thus it is not necessary to know explicitly the structure of each term in the perturbative expansion, and the transition probabilities are not subject to any detailed balance condition. In contrast to the ``worm'' algorithm, in our algorithm the number of external legs of diagrams is not fixed and also becomes a random variable. With only a minor modification of the algorithm, one can consider either disconnected or connected diagrams.

 In order to illustrate this general idea, we have applied it to study the running of the renormalized coupling constant in the scalar field theory with quartic interaction. With our numerical algorithm we were able to obtain the coefficients of perturbative expansions in powers of the bare coupling constant up to 15th order. Resulting series were then resummed using a specially adapted Pade-Borel-Leroy resummation procedure. We have confirmed that the coupling constant approaches zero in the continuum limit in four and five space-time dimensions, and grows in three space-time dimensions. Performance of our algorithm in terms of computer time is comparable to the reported performance of the ``worm'' algorithm for the same theory \cite{Wolff:09:1}. Let us also note that the algorithm of \cite{Wolff:09:1} significantly slows down in the weak-coupling limit $\lambda_0 \rightarrow 0$, while the precision of our algorithm increases in this limit. Disadvantages of our method are the need for an external resummation procedure and the small signal-to-noise ratio in the low-momentum region. The latter is analogous to slowing down of the conventional Monte-Carlo simulations with increasing lattice volume.

 The presented algorithm clearly allows for many improvements. For example, instead of expanding the correlators in powers of the bare coupling constant $\lambda_0$, as in (\ref{correlators_prob_redef}), one could try to expand them in the basis of some specially constructed functions of $\lambda_0$ (see e.g. Chapter 16.5 of \cite{KleinertPhi4}). As well, one can consider other choices of the re-summing coefficients $c_{n,m}$ in (\ref{correlators_prob_redef}) than (\ref{coeff_def}). For example, $c_{n,m}$ can be proportional to the coefficients of the expansion of the averages $\vev{\phi^{n}}$ in powers of the coupling constant $\lambda_0$ in a zero-dimensional scalar field theory with the action $S\lr{\phi} = \phi^2/2 + \lambda_0 \, \phi^4/4$. Preliminary calculations show that such choice allows to explore even higher orders of perturbation theory, although the resummation procedure becomes somewhat more involved. It is also possible to speed up the simulations by analytically calculating the expansion coefficients $G_m\lr{p_1, \ldots, p_n}$ for some small $m$. In particular, if the expansion coefficient $G_m\lr{p_1, p_2}$ for the two-point function is known for some finite $m = m_0$, one can modify Algorithm \ref{alg:rp_phi4} by starting at $m = m_0$ and generating the first momenta in the sequence $\lrc{p_1, p_2}$ with the probability distribution proportional to $G_m\lr{p_1, p_2}$. Since $m$ always increases in Algorithm \ref{alg:rp_phi4}, only diagrams of order $m > m_0$ will be sampled with such a modification. With small changes, the method can be also applied to quantum field theories in the large-$N$ limit \cite{Buividovich:10:2}. In this case the perturbative series are expected to diverge only exponentially, and by analyzing the behavior of large-order expansion coefficients one can locate phase transitions of Gross-Witten type. Since the aim of the present paper is to illustrate the general method on the simplest possible nontrivial example, we do not consider here these potential improvements.

 Let us also discuss briefly the application of Algorithm \ref{alg:rp_phi4} to the $\phi^4$ theory with spontaneous symmetry breaking. A detailed study of this case will be presented elsewhere. Since the possibility to interpret the bare propagator as the probability distribution of the momentum is absolutely crucial for the formulation of Algorithm \ref{alg:rp_phi4}, it is not possible to set $m_0^2 < 0$ without modifying the algorithm. Instead, one should redefine the field variable $\phi\lr{x} = \tilde{\phi}\lr{x} + \phi_0$, where $\phi_0$ corresponds to one of the stable minima of the potential, and write Schwinger-Dyson equations in terms of the field $\tilde{\phi}\lr{x}$, which now has the physical mass. Note that as long as $Z_2$ symmetry remains spontaneously broken, the system will not be able to jump to other nontrivial minimum of the potential (as it would typically happen with the standard Monte-Carlo in a finite volume). The reason is that our method in fact operates in an infinite volume limit, and the infrared cutoff is imposed only when collecting the statistics (see Section \ref{sec:series_resummation}).

 We hope that the presented approach can be easily generalized to other quantum field theories. It can be especially advantageous for those theories, for which the all-order perturbative expansion is difficult to construct in an explicit form or leads to asymptotic series, and hence the ``worm'' algorithm cannot be applied in a straightforward way.

 Of primary interest is the extension to non-Abelian gauge theories with fermions, where Diagrammatic Monte-Carlo can potentially help to tackle the notorious sign problem at finite chemical potential (see, e.g., \cite{DeForcrand:10:1}). In non-Abelian gauge theories Schwinger-Dyson equations can be written in terms of gauge-invariant quantities -- Wilson loops. Such formulation of Schwinger-Dyson equations is known as Migdal-Makeenko loop equations \cite{Migdal:81:1}. The configuration space of the random process which solves these equations should contain loops (that is, closed sequences of links) on the lattice \footnote{It is also possible to formulate the Migdal-Makeenko loop equations for loops in momentum space \cite{Migdal:81:1}, although it is not clear how to construct the appropriate ``functional fourier transform'' in lattice gauge theory.}, and the basic transformations on this space should be the merging and the modification of loops (for examples of such loop transformations see the algorithm for solving the Weingarten model, described in \cite{Buividovich:10:2}). The probability that such random process produces some loop $C$ should be then proportional to the Wilson loop $W\lr{C}$. The main problem with such approach is that straightforward stochastic interpretation of loop equations on the lattice leads to the strong-coupling expansion \cite{Buividovich:10:2}, which is not analytically connected with the continuum limit of the theory at weak coupling. Let us mention, however, that some attempts to explore the phase structure of QCD at finite chemical potential basing on the lowest-order strong-coupling expansion have been already reported in the literature, and quite interesting results were obtained \cite{DeForcrand:10:1}. Work on the application of the presented method to loop equations in non-Abelian gauge theories is in progress.

\begin{acknowledgments}
 The author is grateful to  M. I. Polikarpov, Yu. M. Makeenko, A. S. Gorsky and N. V. Prokof'ev for interesting and stimulating discussions. This work was partly supported by Grants RFBR 09-02-00338-a, RFBR 11-02-01227-a, a grant for the leading scientific schools No. NSh-6260.2010.2, by the Federal Special-Purpose Programme 'Personnel' of the Russian Ministry of Science and Education and by a personal grant from the FAIR-Russia Research Center (FRRC).
\end{acknowledgments}


\fussy
\raggedright



\end{document}